\documentclass[a4paper,11pt]{article}
\pdfoutput=1
 \usepackage{jheppub} 
\usepackage{slashed}

\newcommand{\col}{\textcolor{magenta}}

\def\beq{\begin{equation}}
\def\eeq{\end{equation}}
\def\bea{\begin{eqnarray}}
\def\eea{\end{eqnarray}}
\def\bit{\begin{itemize}}
\def\eit{\end{itemize}}

\def\baa{\begin{array}}
\def\eaa{\end{array}}

\begin{document}
\title{\boldmath Disambiguating Seesaw Models using Invariant Mass Variables at Hadron Colliders}

\author[a,b]{P. S. Bhupal Dev,\note{{\tt  bhupal.dev@tum.de}}}
\author[c]{Doojin Kim,\note{{\tt immworry@ufl.edu}}}
\author[d]{Rabindra N. Mohapatra\note{{\tt rmohapat@umd.edu}}}

\affiliation[a]{Consortium for Fundamental Physics, School of Physics and Astronomy,  \\
University of Manchester, Manchester M13 9PL, United Kingdom}
\affiliation[b]{Physik-Department T30d, Technische Univertit\"{a}t M\"{u}nchen, \\
James-Franck-Stra\ss e 1, 85748 Garching, Germany}
\affiliation[c]{Department of Physics, University of Florida, Gainesville, FL 32611, USA}
\affiliation[d]{Maryland Center for Fundamental Physics and Department of Physics, \\ University of Maryland, College Park, Maryland 20742, USA}

\preprint{\\ TUM-HEP-1022/15, CETUP2015-026, UMD-PP-015-014}

\abstract{We propose ways to  distinguish between different mechanisms behind the collider signals of TeV-scale seesaw models for neutrino masses using kinematic endpoints of invariant mass variables. We particularly focus on two classes of such models widely discussed in literature: (i) Standard Model extended by the addition of singlet neutrinos and (ii) Left-Right Symmetric Models. Relevant scenarios involving the {\em same} ``smoking-gun'' collider signature of dilepton plus dijet with no missing transverse energy differ from one another by their event topology, resulting in distinctive relationships among the kinematic endpoints to be used for discerning them at hadron colliders. These kinematic endpoints are readily translated to the mass parameters of the on-shell particles through simple analytic expressions which can be used for measuring the masses of the new particles. A Monte Carlo simulation with detector effects is conducted to test the viability of the proposed strategy in a realistic environment. Finally, we discuss the future prospects of testing these scenarios at the $\sqrt s=14$ and 100 TeV hadron colliders. }


\maketitle

\section{Introduction}
\label{sec:introduction}
Despite the spectacular experimental progress in the past two decades in determining the  neutrino oscillation parameters~\cite{PDG}, the nature of new physics beyond the Standard Model (SM) responsible for nonzero neutrino masses is still unknown. A widely discussed paradigm for neutrino masses is the so-called  type-I seesaw mechanism~\cite{Minkowski:1977sc,Mohapatra:1979ia,Yanagida:1979as,GellMann:1980vs,Glashow:1979nm} which postulates the existence of heavy right-handed (RH) neutrinos with Majorana masses. The masses of the RH neutrinos, synonymous with the seesaw scale, is {\em a priori} unknown, and its determination would play a big role in vindicating the seesaw mechanism as the new physics responsible for neutrino mass generation. There are a variety of opinions as to where this scale could be~\cite{Mohapatra:2006gs}, ranging from the left-handed (LH) neutrino mass scale of sub-eV all the way up to the grand unification theory (GUT) scale. The GUT approach, which is quite attractive, is mainly motivated by the straightforward embedding of the seesaw mechanism in $SO(10)$ GUTs~\cite{Mohapatra:1979ia}. The seesaw scale in simple $SO(10)$ models is near $10^{14}$ GeV~\cite{Mohapatra:2006gs}  making it quite hard to test in any foreseeable laboratory experiments.\footnote{However, in $SO(10)$ models with multiple intermediate symmetry breaking scales and/or variants of the type-I seesaw mechanism, the seesaw scale can be near a TeV; see~\cite{Malinsky:2005bi, Dev:2009aw, Dev:2010he, Borah:2010kk, Romeri:2011ie, Arbelaez:2013hr, Awasthi:2013ff, Deppisch:2014zta, Parida:2014dla, Dev:2015pga, Deppisch:2015cua, Bandyopadhyay:2015fka} for an incomplete set of examples.} There are also arguments based on naturalness of the Higgs mass which suggest the seesaw scale to be  below $10^7$ GeV or so~\cite{Vissani:1997ys,Clarke:2015hta}. It is therefore of interest to focus on the seesaw scale being in the TeV range since we have the Large Hadron Collider (LHC) searching for many kinds of TeV-scale new physics, which could also probe TeV-scale seesaw through the ``smoking gun'' lepton number violating (LNV) signal of same-sign dilepton plus dijet final states $\ell^\pm \ell^\pm jj$~\cite{Keung:1983uu} and other related processes~\cite{Datta:1993nm, Dev:2013wba, Alva:2014gxa}. In addition,  there are many kinds of complementary low energy searches for rare processes, such as neutrinoless double beta decay ($0\nu\beta\beta$)~\cite{Rodejohann:2011mu}, lepton flavor violation (LFV)~\cite{deGouvea:2013zba}, anomalous Higgs decays~\cite{Dev:2012zg, Cely:2012bz, Maiezza:2015lza, Dermisek:2015vra} and so on which are sensitive to TeV-scale models of neutrino mass. For reviews on various phenomenological aspects of TeV-scale seesaw, see e.g.,~\cite{Atre:2009rg, Drewes:2013gca, Deppisch:2015qwa, Alekhin:2015byh}. 

On the theory front, a natural framework which could provide a TeV-scale renormalizable theory of the seesaw mechanism is the Left-Right (L-R) Symmetric extension of the SM (LRSM)~\cite{Pati:1974yy, Mohapatra:1974hk, Mohapatra:1974gc, Senjanovic:1975rk}. The two essential ingredients of seesaw, i.e., the existence of the RH neutrinos (and three of them) and the seesaw scale,  emerge naturally in LRSM -- the former as the parity gauge partners of the LH neutrinos and the latter as the scale of parity restoration.\footnote{There exist examples~\cite{Dev:2013oxa} where the small neutrino masses via type-I seesaw at TeV-scale can arise without excessive fine tuning of the LRSM parameters.} This model predicts the existence of new gauge bosons $W_R$ and $Z'$ which are the RH counterparts of the familiar SM $W$ and $Z$ bosons, and can be searched for in laboratory experiments. There are therefore considerable theoretical motivations to search for TeV-scale seesaw signatures at the LHC, which could lead to parity restoration. Apart from the spectacular LNV signal of $\ell^\pm \ell^\pm jj$ at the LHC~\cite{Keung:1983uu, Ferrari:2000sp, Nemevsek:2011hz, Chen:2011hc, Chakrabortty:2012pp, Das:2012ii, Saavedra:2012gf, Chen:2013fna, Rizzo:2014xma, Ng:2015hba}, a TeV-scale LRSM also leads to potentially large contributions to the low-energy LNV process of $0\nu\beta\beta$~\cite{Mohapatra:1979ia, Mohapatra:1981pm, Picciotto:1982qe, Hirsch:1996qw, Tello:2010am, Chakrabortty:2012mh, Nemevsek:2012iq, Barry:2013xxa, Dev:2013vxa, Huang:2013kma, Dev:2014xea, Borah:2015ufa, Ge:2015yqa, Awasthi:2015ota}. In addition, there are a plethora of LFV processes, such as $\mu\to e\gamma$, $\mu\to 3e$, and $\mu\to e$ conversion in nuclei, which can also get sizable contributions from the RH sector~\cite{Riazuddin:1981hz, Pal:1983bf, Mohapatra:1992uu, Cirigliano:2004mv, Cirigliano:2004tc, Tello:2010am, Das:2012ii, Barry:2013xxa, Dev:2013oxa, Dev:2014xea, Awasthi:2015ota}.  

As far as the collider signals are concerned, in the case of SM seesaw, the Yukawa part of the model Lagrangian is given by
\bea
{\cal L}^I_{Y} \ = \ {\cal L}_{Y}^{\rm SM}+h_\nu \overline{N}H\psi_L+M_NN^{\sf T}CN+ {\rm H.c.},
\label{smy}
\eea
where $\psi_L$, $H$, $N$ denote respectively the SM lepton and Higgs doublets and the RH neutrino singlet fields under $SU(2)_L$, and $C$ denotes the charge conjugation matrix. We have dropped the generation indices.\footnote{As in most of the collider studies of seesaw, we will assume later a single heavy neutrino giving the dominant contribution to the $\ell\ell jj$ signal for simplicity.}  When $H$ picks up a vacuum expectation value $v$ to break the electroweak gauge symmetry, it leads to the usual type-I seesaw~\cite{Minkowski:1977sc,Mohapatra:1979ia,Yanagida:1979as,GellMann:1980vs,Glashow:1979nm}. On the other hand, in the LRSM, based on the gauge group $SU(2)_L\times SU(2)_R\times U(1)_{B-L}$, we have the additional gauge bosons $W_R,~Z'$ which play an important role in collider phenomenology, if $W_R$ mass is in the accessible range. The part of the Lagrangian relevant to our discussion is that involving the charged current and Yukawa interactions responsible for the seesaw matrix after $SU(2)_R\times U(1)_{B-L}$ symmetry is broken: 
\bea
{\cal L}^{\rm LR} \ \supset \ {\cal L}^I_{Y}+ \frac{g_R}{\sqrt{2}}(\overline{N}\gamma^\mu \ell_R W^+_{R,\mu}+{\rm H.c.}) \, ,
\label{lry}
\eea
where we have again omitted the flavor structure in the RH sector. Note that there is now a direct coupling of the RH neutrinos to the $W_R$ gauge boson which will distinguish its production mode at the LHC from that of the SM seesaw case. Assuming $m_N<m_{W_R}$, as suggested from vacuum stability arguments~\cite{Mohapatra:1986pj}, the basic collider signal arises from the on-shell production of the RH neutrinos accompanied by a charged lepton in the first stage, and the  subsequent decay of $N$ to $\ell jj$ final states (with $\ell=e,\mu,\tau$, depending on the initial flavor of $N$). The latter can go via the virtual intermediate state $W_R$ or physical on-shell $W$, assuming that $m_N > m_{W}$.\footnote{For $m_N<m_{W}$, one can look for other interesting signals like displaced vertices at the LHC~\cite{Helo:2013esa} or decay products of charged mesons~\cite{Felisola:2015bha, Cvetic:2015naa, Dib:2015oka}, which we do not discuss here.} 

It is worth emphasizing that in the LRSM, the Majorana nature of the RH neutrinos inevitably leads to the LNV signature of $\ell^\pm \ell^\pm jj$~\cite{Keung:1983uu}, irrespective of the Dirac Yukawa couplings in Eq.~\eqref{lry}. On the other hand, in the case of the SM seesaw, the strength of the dilepton signal crucially depends on the size of the heavy-light neutrino mixing parameters $V_{\ell N}\sim vh_\nu m_N^{-1}$, $m_N$ being the corresponding eigenvalues of the RH neutrino mass matrix $M_N$ in Eq.~\eqref{smy}. In fact, the constraints of small neutrino masses from type-I seesaw have severe implications for the LNV $\ell^\pm \ell^\pm jj$ signals~\cite{Kersten:2007vk, Ibarra:2010xw, Pavon:2015cga}. Whether the dilepton signal can be of same-sign or mostly of opposite-sign depends on how degenerate the RH neutrinos are and to what extent they satisfy the coherence condition~\cite{Akhmedov:2007fk}.\footnote{For instance, if the pattern of RH neutrino masses is hierarchical while maintaining large $|V_{\ell N}|^2$~\cite{Pilaftsis:1991ug} and  satisfying the neutrino oscillation data,  one can in principle have a $\ell^\pm \ell^\pm jj$ signal~\cite{Datta:1993nm,Dev:2013wba, Alva:2014gxa, Panella:2001wq, Han:2006ip, Aguila:2007em, Aguila:2008cj}.} Nevertheless, the general kinematic strategy presented in this paper is equally applicable to both same and opposite-sign dilepton signals. In this sense, its efficacy is not just limited to the type-I seesaw models, but also to many of its variants, such as the inverse~\cite{Mohapatra:1986aw, Mohapatra:1986bd}, linear~\cite{Akhmedov:1995ip, Barr:2003nn, Malinsky:2005bi} and generalized~\cite{Gavela:2009cd, Dev:2012sg, Dev:2015pga} seesaw models, which typically predict a dominant opposite-sign dilepton signal. In fact, in light of the recent observations from both CMS~\cite{Khachatryan:2014dka} and ATLAS~\cite{Aad:2015xaa} indicating a paucity of $\ell^\pm \ell^\pm jj$ events, the $\ell^\pm \ell^\mp jj$ signal might turn out to be more relevant~\cite{Dev:2015pga, Gluza:2015goa, Dobrescu:2015qna, Deppisch:2015cua} in the potential discovery of parity restoration in future LHC data.  Therefore, we will not specify the lepton charge in our subsequent discussion, and the signal will be simply referred to as the $\ell\ell jj$ signal.  

\begin{figure}[t!]
\centering
\includegraphics[width=7cm]{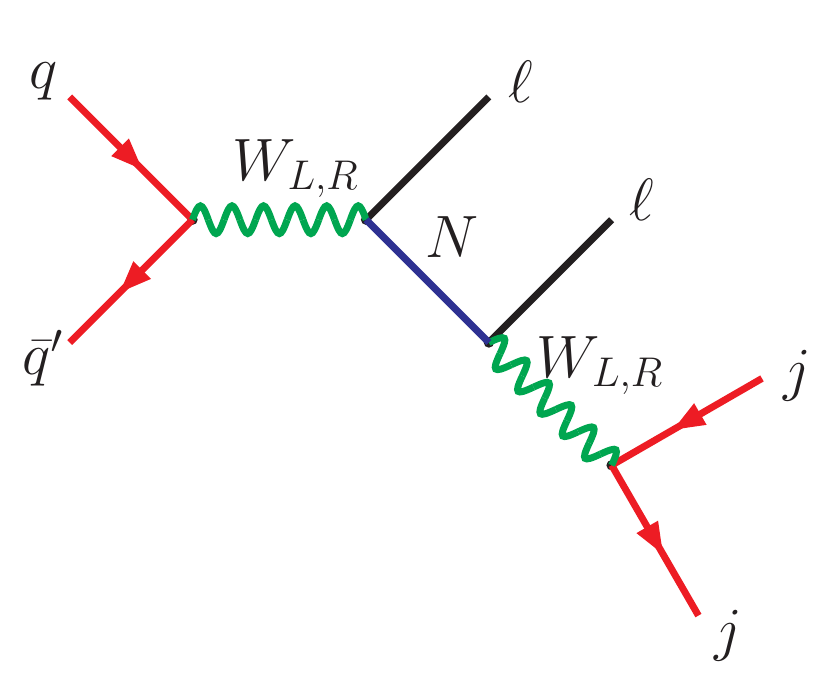}
\caption{Feynman diagrams for the ``smoking gun'' collider signal of seesaw in the LRSM.}
\label{fig1}
\end{figure}

For $m_{W_R}>m_N>m_W$, there are {\em four} different sources in the LRSM for the origin of the $\ell \ell jj$ signal at the LHC~\cite{Chen:2013fna, Dev:2013vba} (see Figure~\ref{fig1}):
\bea
LL:&& \; pp\  \rightarrow \ W_L^* \ \rightarrow \ \ell N \ \rightarrow \ \ell \ell W_L \ \rightarrow \ \ell \ell j j \, ,\label{eq:LL}\\
RR:&& \; pp \ \rightarrow \ W_R \ \rightarrow \ \ell N \ \rightarrow \ \ell \ell W_R^* \ \rightarrow \ \ell \ell j j \, ,\label{eq:RR}\\
RL:&& \; pp \ \rightarrow \ W_R \ \rightarrow \ \ell N \ \rightarrow \ \ell \ell W_L \ \rightarrow \ \ell \ell j j \, ,\label{eq:RL}\\
LR:&& \; pp \ \rightarrow \ W_L^* \ \rightarrow \ \ell N \ \rightarrow \ \ell \ell W_R^* \  \rightarrow \  \ell \ell j j \, ,\label{eq:LR}
\eea
where the first ($LL$) mode is the only one that arises in the SM seesaw via $s$-channel exchange of the SM $W$-boson (which is denoted by $W_L$ to justify the name $LL$), whereas all the four modes can arise in L-R models. These signals are uniquely suited to probe the Majorana and Dirac flavor structure of the neutrino seesaw and are therefore an important probe of the detailed nature of the seesaw mechanism. To this end, it is important to know how to disambiguate the different mechanisms \eqref{eq:LL}-\eqref{eq:LR}. The main point of this paper is to show that determining the kinematic endpoints of different invariant mass distributions,\footnote{For a systematic derivation of the kinematic endpoint and the shape in invariant mass distributions, see, for example, Ref.~\cite{Miller:2005zp}.} e.g., $m_{\ell\ell}$, $m_{\ell jj}$ etc., provides a unique way to distinguish these scenarios, irrespective of the dynamical details. 
Note that given possible scenarios listed above, the invariant mass variables involving a single lepton suffer from the combinatorial ambiguity due to the unawareness of the order between the two leptons. A special prescription is taken for those variables. We then provide a systematic way of determining the kinematic endpoints of the modified variables resulting from such a prescription.
 
Various ways of distinguishing the underlying new physics scenarios by making use of the sharp features of kinematic variables have been studied in several recent works, e.g., in~\cite{Agashe:2010gt, Agashe:2010tu, Cho:2012er, Agashe:2012fs} in the context of dark matter stabilization symmetries. 
In particular, kinematic endpoints are {\it robust} against detailed dynamics of the underlying physics such as spin correlation, i.e., specific and extreme {\it kinematic} configurations are relevant to the kinematic endpoints without the need to know the full dynamical details of the process. Moreover, they are typically protected from event selection cuts unless such cuts are customized to reject the events corresponding to the kinematic endpoints. These observations motivate us to utilize the endpoints of various invariant mass variables to distinguish between the seesaw signals \eqref{eq:LL}-\eqref{eq:LR}. We emphasize that different scenarios typically give rise to different event topologies, resulting in the kinematic endpoints having distinctive dependencies upon underlying mass parameters. The identification of specific relationships among them will enable us to uncover the relevant model parameter space and may eventually lead to a measurement of the masses of the involved heavy particles. Hence, we argue that this method is more robust than the dynamical variables like spin or angular correlations, as proposed earlier to distinguish the seesaw signals~\cite{Chen:2013fna, Han:2012vk}. 

The rest of the paper is organized as follows. We first define our notations to be used in subsequent sections, followed by a discussion about the general strategy, in Section~\ref{sec:notation}. We then give  detailed derivations for the kinematic endpoints of various invariant mass variables case-by-case in Section~\ref{sec:derivation}. In Section~\ref{sec:topology}, we discuss various ways for the topology disambiguation and mass measurements of the new particles. In Section~\ref{sec:app}, we illustrate the utility of the endpoint technique for the L-R seesaw signals by performing a Monte Carlo simulation including detector effects. In Section~\ref{sec:future}, we discuss the L-R seesaw phase diagram for collider studies and the future prospects of probing the LRSM parameter space at the $\sqrt s=14$ TeV LHC, as well as its complementarity with other low-energy probes. We also derive the first LRSM sensitivity contours for the planned $\sqrt s=100$ TeV $pp$ collider.  Our conclusions are given in Section~\ref{sec:conclusions}. In Appendix~\ref{sec:AppA}, we illustrate some invariant mass distributions for the LRSM. 

\section{Notations and general strategy \label{sec:notation}}
\begin{figure}[t]
\centering
\includegraphics[width=15cm,height=5.5cm]{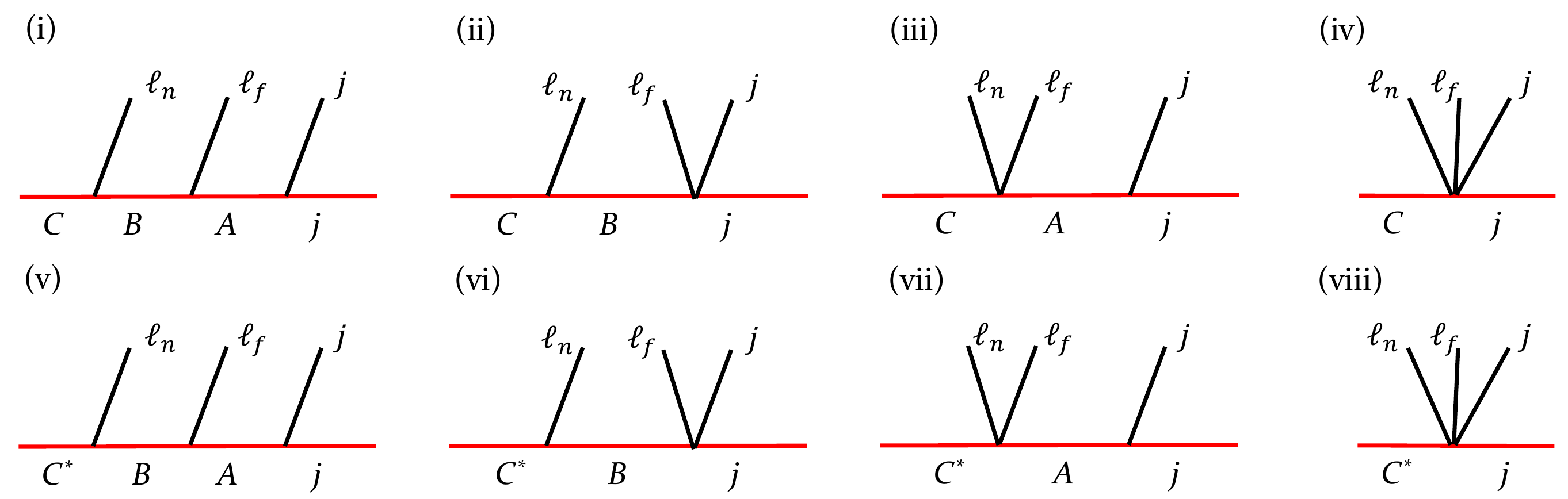}
\caption{\label{fig:decaytopo} Eight possible event topologies for the $\ell\ell jj$ final state. $C^*$ in the second row implies that it is off-shell.}
\end{figure}

To accommodate all possibilities for an $\ell \ell jj$ final state in a model-independent manner, we  generalize the scenarios introduced in Eqs.~\eqref{eq:LL}-\eqref{eq:LR} according to their decay topologies in conjunction with symbolic notations. We begin with a three-step cascade decay sequence of a heavy resonance $C$:
\bea
C \ \rightarrow \ \ell_n B  \ \rightarrow \ \ell_n \ell_f A \ \rightarrow \ \ell_n \ell_f j j \, ,
\eea
where $\ell_n$ and $\ell_f$ are correspondingly identified as ``near'' and ``far''-lepton with respect to the particle $C$ to identify their relative location to each other. The associated event topology is explicitly diagrammed in Figure~\ref{fig:decaytopo}(i), being henceforth denoted as Case (i). If particle $A$ is heavier than particle $B$, then the latter directly decays into $\ell_f jj$ via a 3-body decay as shown in Figure~\ref{fig:decaytopo}(ii), which is denoted as Case (ii). In an analogous manner, one can imagine the situation where particle $B$ is heavier than particle $C$ so that the latter decays into two leptons and particle $A$ via a 3-body decay. The relevant diagram is shown in Figure~\ref{fig:decaytopo}(iii) and denoted as Case (iii). Another possibility is the situation where particle $C$ directly decays into two leptons and two jets via a 4-body decay, as shown in Figure~\ref{fig:decaytopo}(iv) and labelled as Case (iv). Finally, we consider the situations where particle $C$ is off-shell (denoted as $C^*$) for Cases (i) through (iv), and their counterparts are respectively exhibited in Figure~\ref{fig:decaytopo}(v) to~\ref{fig:decaytopo}(viii), and labelled as Cases (v) to (viii). Note that Case (viii) is a very unlikely scenario, as $C$ is always assumed to be heavier than the $\ell\ell jj$ system,  so we do not consider it any further. 

\begin{table}[t!]
\centering
\begin{tabular}{c|c|cc}\hline\hline
Topology & Region(s) & \multicolumn{2}{c}{Model Scenario(s)} \\
\hline \hline
(i) & $\mathcal{R}_{1(1)}$, $\mathcal{R}_{1(2)}$ & \col{$R_lL_h$} & \col{($m_{W_R}>m_N>m_{W}$)} \\ \hline
&   & \col{$R_lR_l$} & \col{($m_{W_R}>m_N$)}\\ 
(ii) & $\mathcal{R}_{2(1)}$, $\mathcal{R}_{2(2)}$ & $R_lL_l$ & ($m_{W_R}>m_{W}>m_N$)\\
& & $L_lR_l$ & ($m_{W_R}>m_{W}>m_N$)\\
& & $L_lL_l$ & ($m_{W}>m_N$)\\ \hline
(iii) & $\mathcal{R}_{3}$ & $R_hL_h$ & ($m_N>m_{W_R}>m_{W}$) \\ \hline 
 & & $R_h R_h$ & ($m_N > m_{W_R}$) \\
(iv) & $\mathcal{R}_{4}$ & $L_hR_l$ & ($m_N>m_{W_R}>m_{W}$) \\ 
 & & $L_h R_h$ & ($m_N > m_{W_R} >m_W$) \\ 
 & & $L_h L_h$ & ($m_N > m_W$) \\ \hline 
&   & $R_hR_h$ & ($m_N>m_{W_R}$) \\ 
(v) & $\mathcal{R}_{5}$ & $R_hL_h$ & ($m_N>m_{W_R}>m_{W}$) \\
& & $L_hR_h$ & ($m_N>m_{W_R}>m_{W}$) \\
& & \col{$L_hL_h$} & \col{($m_N>m_{W}$)} \\ \hline
(vi) & $\mathcal{R}_{6}$ & \col{$L_hR_l$} & \col{($m_{W_R}>m_N>m_{W}$)} \\ \hline
(vii) & $\mathcal{R}_{7}$ & $L_h R_l$ &  ($m_{W_R}>m_N>m_{W}$) \\
\hline \hline
\end{tabular}
\vspace{0.5cm}
\caption{\label{tab:assignment} Relations of the event topologies (i)--(vii) shown in Figure~\ref{fig:decaytopo} with the relevant kinematic regions labelled as $\mathcal{R}_1-\mathcal{R}_7$, as well as their mapping with the possible scenarios in the LRSM. The subregions in parenthesis, e.g.,  $\mathcal{R}_{1(1)}$ and $\mathcal{R}_{1(2)}$,  mean that they correspond to the same event topology but with different kinematic endpoints, as explained later in Section~\ref{sec:derivation} [cf.~Eqs.~\eqref{eq:caseiregion}, \eqref{mljj} and \eqref{mlj2}]. }
\end{table}

Provided with the L-R seesaw models described in the previous section, one can easily correlate these event topologies with various scenarios arising therein by identifying $N$ as particle $B$ and various combinations of gauge bosons ($W_L$, $W_L$), ($W_R$, $W_R$), ($W_R$, $W_L$), and ($W_L$, $W_R$) as particles ($C$, $A$), respectively, depending on the underlying scenario. To accommodate all possible mass hierarchies, we introduce a subscript $h$ ($l$) meaning that $N$ is heavier (lighter) than the subscripted gauge boson, e.g., $R_lL_h$ means $m_{W_R}>m_N>m_{W}$. For $LL$, $RR$, $RL$, and $LR$, each letter accepts either $l$ or $h$ subscript so that (naively) 16 scenarios would be possible. However, the subscripts in four scenarios such as $L_lL_h$, $L_hL_l$, $R_lR_h$, and $R_hR_l$ imply a contradictory mass spectrum, i.e., the same gauge boson (either $W_L$ or $W_R$) cannot be heavier and lighter than $N$ simultaneously. In addition, we have only assumed $m_{W_R}>m_{W}$, as required to satisfy the experimental constraints from direct searches at the LHC ~\cite{Aad:2015xaa, Khachatryan:2014dka}, as well as the indirect constraints from $K_L-K_S$ mass difference~\cite{Beall:1981ze, Zhang:2007da, Maiezza:2010ic}. Therefore, the scenarios implying $m_W >m_{W_R}$, i.e., $R_h L_l$ and $L_l R_h$,  are not allowed. As a result, only 10 different combinations are possible. In principle, the particle identified as $C$ can be forced to be on-shell (assuming that it is within the kinematic reach of the collider experiment), so that all 10 possibilities can be assigned to the first four event topologies, listed in Table~\ref{tab:assignment}. Depending on the underlying model details, particle $C$ can also be off-shell, if $m_C< m_B$, or even $N$ can be off-shell. The relevant identifications to the last three topologies are also tabulated in Table~\ref{tab:assignment}.
Note that the scenarios listed in Eqs.~\eqref{eq:LL}-\eqref{eq:LR} can be obtained readily by imposing the additional assumption that $m_{W_R}>m_N>m_{W}$, as shown by the colored text in Table~\ref{tab:assignment}. A detailed topology identification of these scenarios with respect to the kinematic regions ${\cal R}_1$ to ${\cal R}_7$ will be discussed in Sections~\ref{sec:topology} and~\ref{sec:app}.


With the final state of $\ell_n\ell_f jj$, one can come up with eight non-trivial invariant mass variables, namely, $m_{\ell\ell},\; m_{\ell_n j},\; m_{\ell_f j},\; m_{jj},\; m_{\ell\ell j},\; m_{\ell_n jj},\; m_{\ell_f jj},$ and $m_{\ell\ell jj}$, where we omit the subscript on the lepton for the variables involving both leptons. We remark that the two jets are emitted from the same particle for all eight cases, and thus we do not distinguish one from the other. An immediate experimental challenge is that one does not know  {\it a priori} which lepton comes first with respect to the $C$ particle, although we theoretically label them as $\ell_n$ and $\ell_f$. This innate combinatorial issue particularly becomes a major hurdle to form invariant mass variables involving a single lepton. For example, $m_{\ell_n j}$ or $m_{\ell_f j}$ are not valid experimental observables again because $\ell_n$ and $\ell_f$ are not discernible event-by-event. 

Given an invariant mass variable having such a combinatorial ambiguity, a possible prescription to resolve it is to evaluate both invariant mass values for a fixed set of jets and divide them into the bigger and the smaller ones, labelled by the superscripts ``$>$'' and ``$<$'', respectively. Taking the example of the invariant masses formed by a lepton and a jet, we compute $m_{\ell_n j}$ and $m_{\ell_f j}$ for a given $j$ and assign the bigger (smaller) one to $m_{\ell j}^>$ ($m_{\ell j}^<$). A similar rule is applied to the invariant mass variables constructed by a lepton and two jets, yielding $m_{\ell jj}^>$  and $m_{\ell jj}^<$. With this prescription, we end up with the following eight experimentally valid observables:
\bea
\hbox{2-body}&:& m_{\ell\ell},\;m_{jj},\;m_{\ell j}^>,\;m_{\ell j}^< \, , \label{eq:2body} \\
\hbox{3-body}&:& m_{\ell\ell j},\;m_{\ell jj}^>,\;m_{\ell jj}^< \, , \label{eq:3body} \\
\hbox{4-body}&:& m_{\ell\ell jj} \label{eq:4body} \, .
\eea
We basically measure the kinematic endpoints of these eight variables to identify the underlying decay topology by examining their interrelationship and kinematic features which are the main subjects of the next section.\footnote{Of course, the kinematic endpoints for $m_{\ell j}^>$, $m_{\ell j}^<$, and $m_{\ell \ell j}$ are better saturated than those for the others, because the presence of two jets enables us to obtain two values for the three while we get only a single value for the others.}



\section{Derivation of kinematic endpoints \label{sec:derivation}}
In this section we derive the analytic expressions for the kinematic endpoints of the invariant mass variables listed in Eqs.~(\ref{eq:2body})--(\ref{eq:4body}). We first elaborate the detailed steps in obtaining the final expressions when particle $C$ is on-shell, and later we briefly mention how the relevant results can be applicable to the case of off-shell $C$. Our main focus is the kinematic endpoints that are insensitive to the details of relevant decays, and thus we do not hypothesize any specific matrix element, i.e., we deal with only the phase-space structure for the associated decay kinematics. 

\subsection{Case (i): $m_C>m_B>m_A$ }
\begin{figure}[t]
\centering
\includegraphics[scale=0.75]{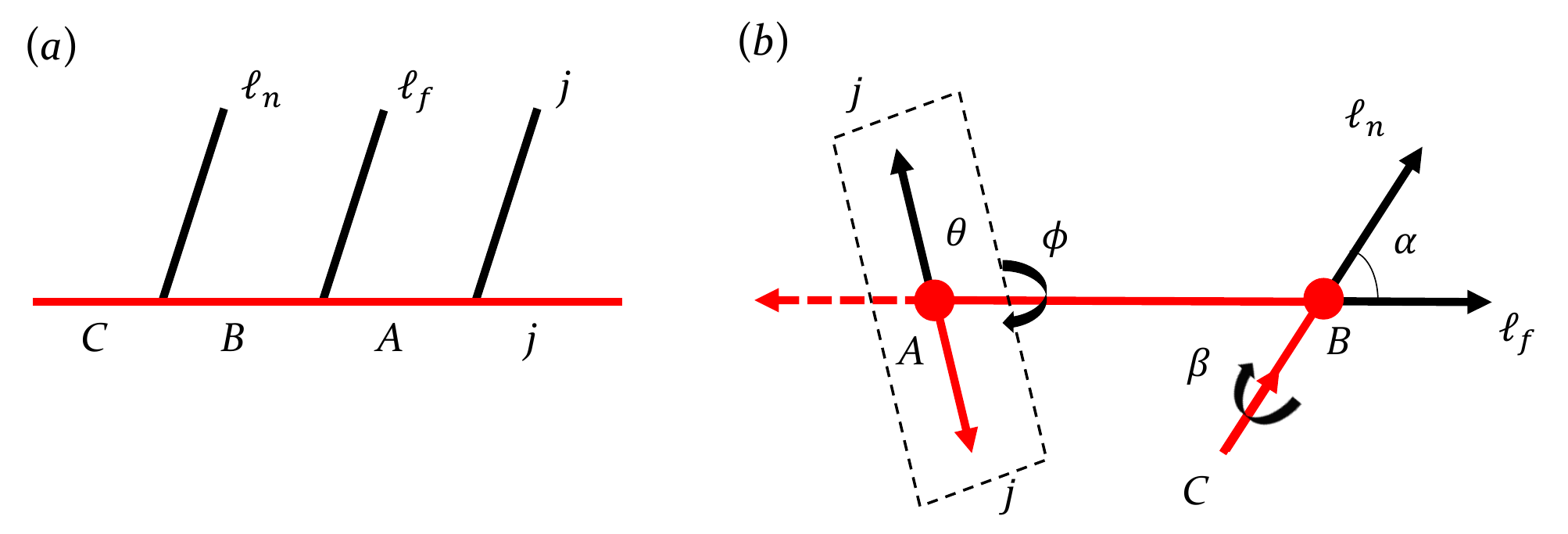}
\caption{\label{fig:RL} The event topology for Case (i) (left panel) and the relevant kinematics described in the rest frame of particle $B$ (right panel). Here, $\alpha$ is the polar angle of $\ell_f$ with respect to the direction of $\ell_n$ (or equivalently, particle $C$) while $\theta$ is the polar angle of $j$ (black arrow) with respect to the direction of $B$ in the rest frame of particle $A$. $\beta$ and $\phi$ are azimuthal angles for $\ell_n$ and $j$ (black arrows) around the horizontal axis defined by $\ell_f$.  }
\end{figure}
This scenario can be represented by a three-step {\it on}-shell cascade decay of a heavy resonance $C$, as shown in Figure~\ref{fig:RL}($a$). We emphasize that it is rather convenient to describe the associated kinematic configuration in the rest frame of particle $B$, as shown in Figure~\ref{fig:RL}($b$).\footnote{\label{ft:1}Although angle $\theta$ is defined with respect to one jet of the two, one could develop the parallel argument with respect to the other jet. Since $\theta$ is a quantity measured in the rest frame of particle $A$, exactly the same argument goes through with $\cos\theta \rightarrow -\cos\theta$. However, this does not provide additional information due to the fact that $\cos\theta$ ($-\cos\theta$) spans $+1$ ($-1$) to $-1$ ($+1$) as $\theta$ increases, and in turn, any similar prescription which would be applied to jet-induced combinatorics for a fixed set of leptons does not enable us to obtain any further independent information. } Naively, the total degrees of freedom are four, but one angle $\beta$ can be dropped by taking into account the azimuthal symmetry of the system. The mass hierarchy in this case determines the baseline inequalities defining the region of interest:
\bea
0\ < \ R_{BC}<1 \quad \hbox{ and } \quad 0 \ < \ R_{AB} \ <1 \  \, ,
\label{casei}
\eea
where $R_{ij}$ denotes the mass ratio between the massive states $i$ and $j$: $R_{ij}\equiv m_i^2/m_j^2$.

\subsubsection*{2-body invariant mass variables}
We begin with the invariant mass of two leptons $m_{\ell\ell}$, for which the kinematic endpoint is well-known:
\bea
\left( m_{\ell\ell}^{\max} \right)^2 \ = \ m_C^2(1-R_{AB})(1-R_{BC}) \ \equiv  \ m_C^2 y_m \, .
\label{mll}
\eea
Trivially, $m^2_{jj}$ is given as a resonance of $m_A^2$:
\bea
\left(m_{jj}^{\rm max}\right)^2 \ = \ m_A^2 \ \equiv \ m_C^2 R_{BC}R_{AB} \, .
\eea
In order to derive the expressions for $m_{\ell j}^{>,\max}$ and $m_{\ell j}^{<,\max}$, we first rewrite the relevant two invariant masses in terms of angular variables introduced in Figure~\ref{fig:RL}($b$):
\bea
\frac{(p_{\ell_f}+p_j)^2}{m_C^2}&\ \equiv \ &  x \ = \ \frac{ x_m}{2} (1-\cos\theta)\, , \label{eq:x} \\
\frac{(p_{\ell_n}+p_j)^2}{m_C^2}& \ \equiv \ & z \ = \ \frac{z_m e^{-\eta}}{2}\lbrace \cosh \eta - \cos \theta \sinh \eta - \cos \alpha (\cos \theta \cosh \eta - \sinh \eta) \nonumber \\
&& \qquad \qquad \qquad \qquad -\cos \phi\sin \theta \sin \alpha \rbrace \, , \label{eq:z}
\eea
with $\eta$, $x_m$, and $z_m$ defined as follows:
\bea
\eta \ \equiv \ \frac{1}{2}\log R_{AB}^{-1}\, , \qquad x_m \ \equiv\  R_{BC}(1-R_{AB})\, ,\qquad z_m \ \equiv \ 1-R_{BC} \, .
\eea
Note that the invariant mass variables in Eqs.~\eqref{eq:x} and \eqref{eq:z} are normalized by $m_{C}^2$ for later convenience.

\begin{figure}[t]
\centering
\includegraphics[width=7cm,height=7cm]{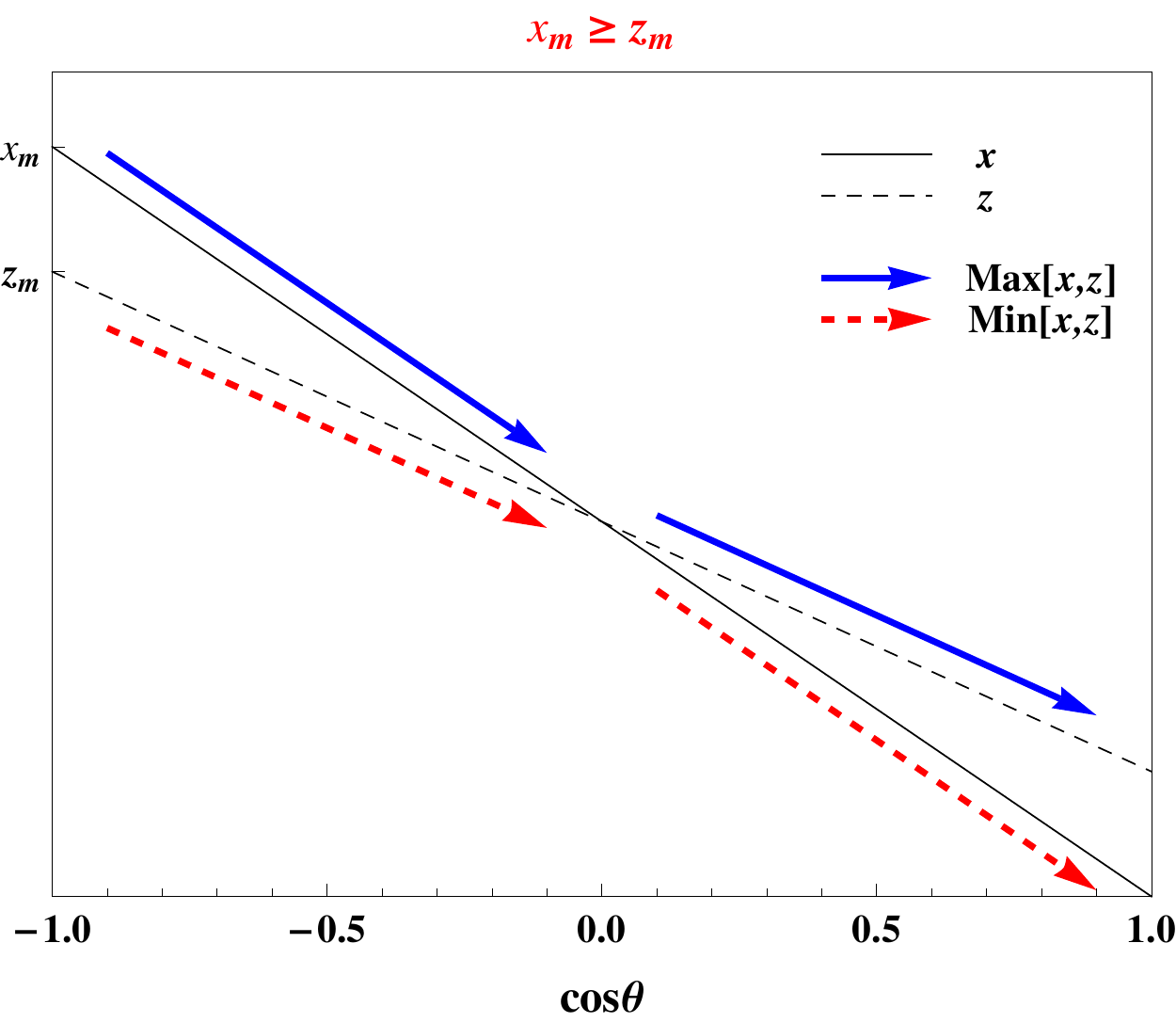}\hspace{0.5cm}
\includegraphics[width=7cm,height=7cm]{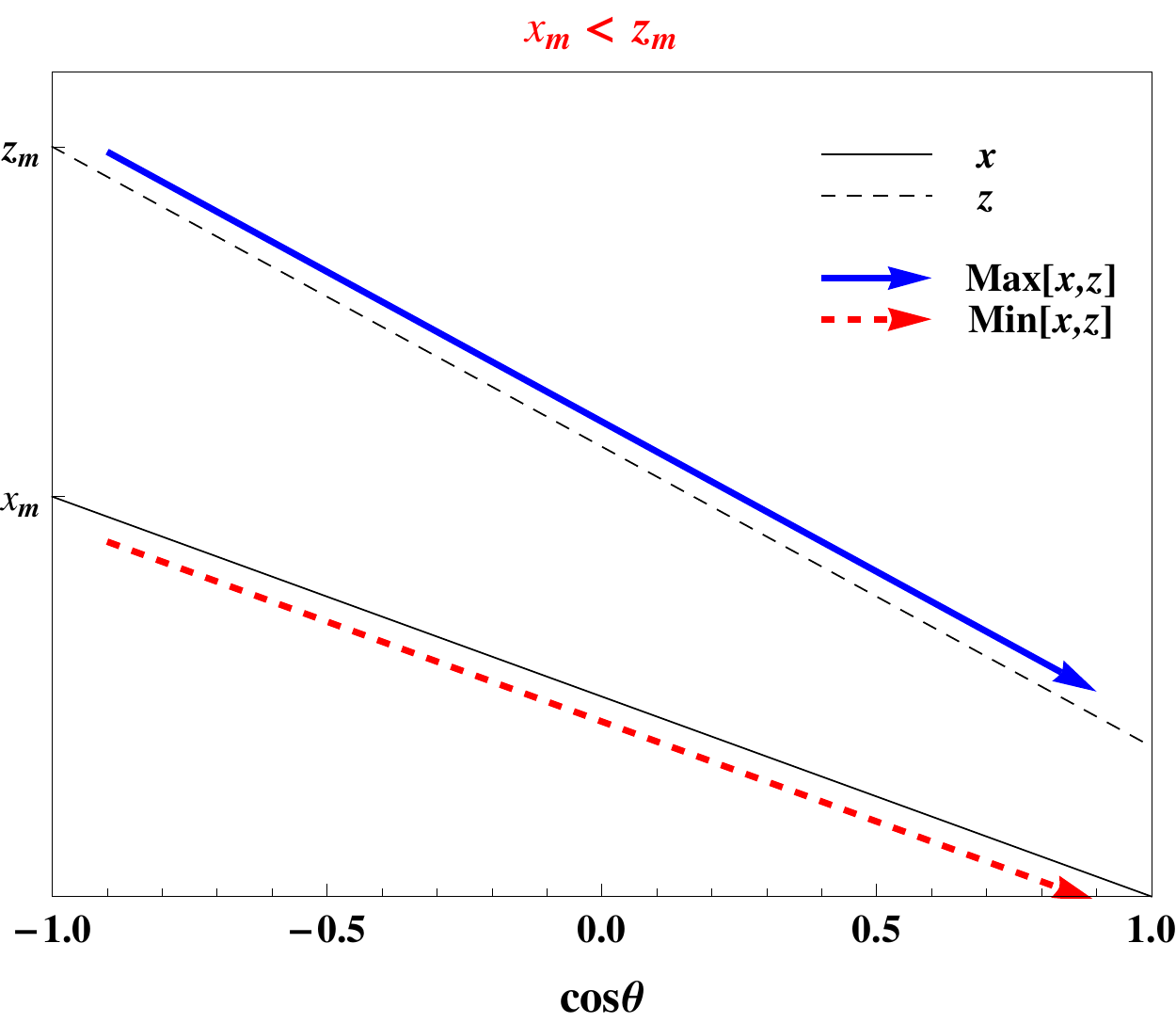}
\caption{\label{fig:InvRLEx} Maximum $x$ and $z$ values in $\cos\theta$ and the way of reading maximum Max[$x,z$] and Min[$x,z$] in $\cos\theta$ for the case of $x_m \geq z_m$ (left panel) and $x_m < z_m$ (right panel).}
\end{figure}

Since the angular variables $\alpha$ and $\phi$ are irrelevant to variable $x$, we first maximize the variable $z$ over them to end up with an expression of $z$ as a function of only $\cos\theta$:
\bea
z&\ = \ &\frac{z_m e^{-\eta}}{2}\left(\cosh \eta -\cos \theta \sinh \eta +\sqrt{1-\cos^2\theta+(\sinh \eta-\cos \theta \cosh \eta)^2}\right) \nonumber \\
& \ = \ & z_m e^{-\eta}(\cosh \eta -\cos \theta \sinh \eta) \, , \label{eq:z}
\eea
which is nothing but a straight line in $\cos\theta$ with a negative slope. To obtain the maxima for $m_{\ell j}^>$ and $m_{\ell j}^<$, we compare Eqs.~(\ref{eq:x}) and~(\ref{eq:z}) while varying $\cos\theta$. Since $x|_{\cos\theta=1}=0$ while $z|_{\cos\theta=1}\neq 0$, two topologically different cases arise as shown in Figure~\ref{fig:InvRLEx}. Each line describes the maximum $x$ or $z$ value for a given $\cos\theta$. For $x_m \geq z_m$ (left panel), one can identify which invariant mass is greater or smaller for any given event, i.e., for a fixed $\cos\theta$. Such a comparison can be conducted for all values of $\cos\theta$, and the trajectories for the greater and the smaller are explicitly delineated by blue-solid and red-dashed arrows respectively. The kinematic endpoint can be obtained simply by reading off the maximum value of each line. For the current example (left panel), they are $x_m$ and $z_m$, and vice versa for $x_m<z_m$ (right panel). Thus, the invariant masses can be formulated as follows:
\bea
\left[ \left(m_{\ell j}^{>,\max}\right)^2,\;\left(m_{\ell j}^{<,\max}\right)^2 \right] \ = \ \left\{
\begin{array}{l}
m_C^2[x_m,\;z_m] \;\;\;\hbox{for }x_m \geq z_m \  \Leftrightarrow \ R_{BC} \in \left[\frac{1}{2-R_{AB}},\;1\right] \cr \cr
m_C^2[z_m,\;x_m] \;\;\;\hbox{for }x_m < z_m \ \Leftrightarrow \ R_{BC} \in \left(0,\; \frac{1}{2-R_{AB}}\right]
\end{array}\right. \label{eq:caseiregion}
\eea
with $R_{AB}\in (0,1]$. We denote the first and the second regions in Eq.~(\ref{eq:caseiregion}) as $\mathcal{R}_{1(1)}$ and $\mathcal{R}_{1(2)}$, respectively, and their link to specific model scenarios are tabulated in Table~\ref{tab:assignment}.

\subsubsection*{3-body invariant mass variables}
For the kinematic endpoint for $m_{\ell\ell j}^2$, we simply use the relevant expression from  Ref.~\cite{Burns:2009zi}:
\bea
\left(m_{\ell\ell j}^{\max}\right)^2 \ = \ m_C^2(1-R_{AB}R_{BC}) \, .
\eea
Considering the other two variables $m_{\ell jj}^>$ and $m_{\ell jj}^<$, we remind the reader that the two jets originate from a common mother particle $A$, and thus they are equivalent to $m_{\ell A}^>$ and $m_{\ell A}^<$, i.e., they are reduced to 2-body invariant mass variables involving a massive visible particle. To obtain their analytic expressions, we first find the formulae for $m_{\ell_f A}^2$ and $m_{\ell_n A}^2$. The former is nothing but the resonance $B$, i.e., $m_{\ell_f A}^2=m_C^2 R_{BC}$. The maximum of the latter can be easily obtained by going through the steps leading to $\left(m_{\ell \ell}^{\max}\right)^2$ [cf.~Eq.~\eqref{mll}]:
\bea
\left(m_{\ell_n A}^{\max}\right)^2 \ = \ m_C^2[1-R_{BC}(1-R_{AB})] \, .
\eea
Since $m_{\ell_f A}^2$ is fixed whatever $m_{\ell_n A}^2$ is given, the kinematic endpoints for the variables of interest can be obtained, as follows: 
\bea
\left[ \left(m_{\ell jj}^{>,\max}\right)^2, \left(m_{\ell jj}^{<,\max}\right)^2 \right]
\ = \ \left\{
\begin{array}{l}
m_C^2\left[R_{BC},\;1-R_{BC}(1-R_{AB})\right] \;\hbox{for }R_{BC} \in \left[\frac{1}{2-R_{AB}},1\right] \cr \cr
m_C^2\left[1-R_{BC}(1-R_{AB}),R_{BC}\right] \;\hbox{for }R_{BC} \in \left(0, \frac{1}{2-R_{AB}}\right]
\end{array}\right.
\label{mljj}
\eea
where the first and the second regions correspond to $\mathcal{R}_{1(1)}$ and $\mathcal{R}_{1(2)}$, respectively.

\subsubsection*{4-body invariant mass variable} 
The 4-body invariant mass variable $m_{\ell\ell jj}$ is trivially given by the mass of resonance $C$: $m_{\ell\ell jj}^2=m_C^2$, since $C$ is on-shell for Cases (i)--(iv).

\subsubsection*{Applying the formulae to Case (v)}
Once the particle $C$ is off-shell, as in Case (v) shown in Figure~\ref{fig:decaytopo}, one can interpret that its mass is given by the center of mass energy $\sqrt{\hat{s}}$. Therefore, it is possible to reuse all analytic expressions thus far by replacing $m_C^2$ by $\hat{s}_{\max}=s$. The chance of reaching $\hat{s}_{\max}$ is, however, so small that any associated endpoints depending on $s$ are not typically saturated. Therefore, only the variables having no dependence on $s$ are reliable: $m_{jj}^2$, one of $m_{\ell j}^2$'s, and one of $m_{\ell jj}^2$'s. Since $s$ is assumed much larger than the mass of particle $B$, $R_{BC}\rightarrow 0$,~\footnote{To avoid any potential confusion, we remark that the true mass for $C$ is irrelevant for $R_{BC}$ once $C$ is off-shell.} so that the corresponding region (denoted by $\mathcal{R}_{5}$ in Table~\ref{tab:assignment}) is defined by
\bea
R_{BC} \ = \ 0 \hbox{ and } 0 \ < \ R_{AB} \ < \ 1 \, .
\label{casev}
\eea

\subsection{Case (ii): $m_C>m_B$ and $m_A\geq m_B$ }
\begin{figure}[t]
\centering
\includegraphics[scale=0.75]{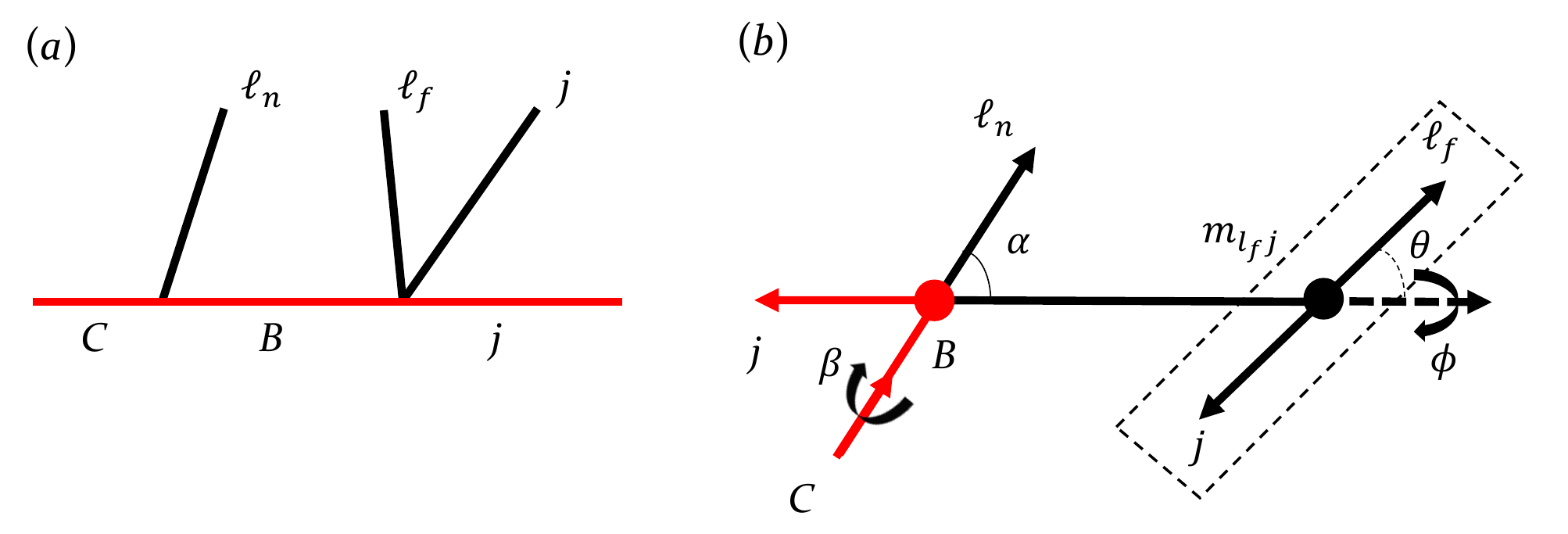}
\caption{\label{fig:RR} The event topology of Case (ii) (left panel) and the relevant kinematics described in the rest frame of particle $B$. Here $\alpha$ is the polar angle of $\ell_n$ with respect to the direction of the composite system of $\ell_f$ and $j$ (black arrow) while $\theta$ is the polar angle of particle $\ell_f$ with respect to the direction of the composite system in the rest frame of particle $B$. $\beta$ and $\phi$ are azimuthal angles for particles $\ell_n$ and $\ell_f$ about the axis extended by $j$ (red arrow) and the composite system.  }
\end{figure}

This scenario can be represented by a two-step cascade decay of a heavy resonance $C$, where the second step is proceeded via a 3-body decay. The corresponding diagram and kinematic configuration are delineated in Figure~\ref{fig:RR}.
It turns out that it is convenient to do the analysis in the rest frame of particle $B$ as shown in Figure~\ref{fig:RR}($b$). Just like Case (i), the angle $\beta$ is irrelevant to the description of the system due to the azimuthal symmetry. The mass hierarchy for this case defines the baseline inequalities determining the corresponding region as
\bea
0 \ < \ R_{BC} \ < \ 1 \quad \hbox{ and } \quad R_{AB} \ \geq  \ 1 \, .
\label{caseii}
\eea

\subsubsection*{2-body invariant mass variables} 
Since particle $B$ decays into three final states via off-shell $A$, $m_{jj}^2$ is given by a distribution, {\it not} a resonance as in Case (i), and the analytic expression for its endpoint is
\bea
\left(m_{jj}^{\max}\right)^2 \ = \ m_C^2R_{BC} \, .
\eea
The kinematic endpoint for $m_{\ell\ell}^2$ can be found in Ref.~\cite{Lester:2006cf}, i.e., 
\bea
\left(m_{\ell\ell}^{\max}\right)^2 \ = \ m_C^2(1-R_{BC}) \, .
\eea
When it comes to the analytic expressions for $m_{\ell j}^{>,\max}$ and $m_{\ell j}^{<,\max}$, we go through a similar argument to that in Case (i). Denoting $(p_{\ell_f}+p_j)^2\equiv m_C^2 x'$, we have
\bea
z =\frac{z_m e^{-\eta}}{2}\lbrace \cosh \eta - \cos \theta \sinh \eta - \cos \alpha (\cos \theta \cosh \eta - \sinh \eta)-\cos \phi\sin \theta \sin \alpha \rbrace
\eea
with $2\eta\equiv \log(R_{BC}/x')$. We observe that $x'_m$ is simply given by $R_{BC}$. Unlike the previous case, $\cos\theta$ is irrelevant to $x'$ so that one can maximize $z$ even over $\cos\theta$ as well as $\phi$ and $\alpha$, yielding $z=z_m$ which is nothing but a horizontal line in $x'$. Clearly, only two cases are available, and the respective expressions are
\bea
\left[ \left(m_{j\ell}^{>,\max}\right)^2,\;\left(m_{j\ell}^{<,\max}\right)^2 \right] \ = \ \left\{
\begin{array}{l}
m_C^2[x'_m,\;z_m] \;\;\;\hbox{for }x'_m \geq z_m \ \Leftrightarrow \ R_{BC} \in \left[\frac{1}{2},\;1\right] \cr \cr
m_C^2[z_m,\;x'_m] \;\;\;\hbox{for }x'_m < z_m \ \Leftrightarrow \ R_{BC} \in \left(0,\; \frac{1}{2}\right],
\end{array}\right.
\label{mlj2}
\eea
and we denote the first and the second regions as $\mathcal{R}_{2(1)}$ and $\mathcal{R}_{2(2)}$, respectively.

\subsubsection*{3-body invariant mass variables} 
To obtain the expression for $\left(m_{\ell\ell j}^{\max}\right)^2$, the result in Ref.~\cite{Lester:2006cf} can be reused, simply yielding $m_C^2$. For the other two variables, we again follow a similar argument to the previous case:
\bea
\left[ \left(m_{\ell jj}^{>,\max}\right)^2,\; \left(m_{\ell jj}^{<,\max}\right)^2 \right] \ = \ m_C^2[1,\; R_{BC}].
\eea


\subsubsection*{Applying the formulae to Case (vi)} 
As in Case (i), we can simply reuse all expressions derived thus far along with replacement of $m_C^2$ by $s$ for an off-shell $C$. Again, the chance of having $\hat{s}_{\max}=s$ is so small that the kinematic endpoints depending on $s$ are not typically saturated. Therefore, only the variables having no dependence on $s$ are reliable: $m_{jj}^2$, one of $m_{\ell j}^2$'s, and $\left(m_{\ell jj}^{<}\right)^2$. As in the previous case, $s$ is assumed to be much larger than the mass of particle $B$, and thus the inequalities defining the corresponding region (denoted by $\mathcal{R}_6$) are given by
\bea
R_{BC} \ = \ 0 \quad \hbox{ and } \quad R_{AB} \ \geq \ 1 \, .
\label{casevi}
\eea
\subsection{Case (iii): $m_B \geq m_C > m_A$}
\begin{figure}[t]
\centering
\includegraphics[scale=0.75]{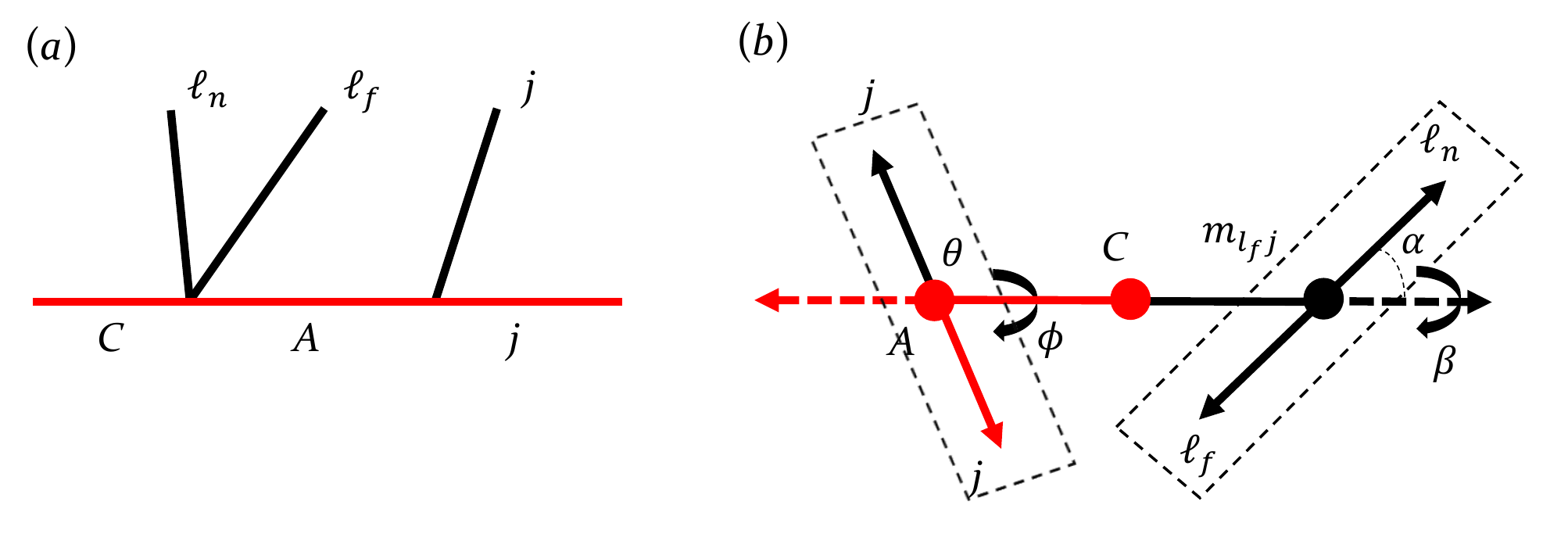}
\caption{\label{fig:RhNL} The event topology of Case (iii) (left panel) and the relevant kinematics described in the rest frame of particle $C$. $\alpha$ is the polar angle of $\ell_n$ with respect to the direction of the composite system of $\ell_n$ and $\ell_f$ while $\theta$ is the polar angle of $j$ (black arrow) with respect to the direction of such a composite system at the rest frame of particle $C$. $\beta$ and $\phi$ are azimuthal angles for $\ell_n$ and $j$ (black arrow) about the axis extended by $j$ (red arrow) and the composite system.  }
\end{figure}

This scenario can be represented by a two-step cascade decay of a heavy resonance $C$, where the first step is proceeded via a 3-body decay. The corresponding diagram and kinematic configuration are delineated in Figure~\ref{fig:RhNL}.
It turns out that it is convenient to do analysis in the rest frame of particle $C$, as shown in Figure~\ref{fig:RhNL}($b$).\footnote{A similar treatment for the other jet is applied as in footnote~\ref{ft:1}.} Just like the previous two cases, azimuthal angle $\beta$ is irrelevant to the description of the system due to the azimuthal symmetry. Though particle $B$ is heavy enough to be integrated out, $R_{AC}$ should be still less than 1, i.e., $R_{AC}=R_{AB}R_{BC}<1$. Therefore, the baseline inequalities defining the corresponding region (denoted by $\mathcal{R}_3$) are
\bea
1\ \leq \ R_{BC}\ < \ \frac{1}{R_{AB}} \quad \hbox{ for } 0 \ < \ R_{AB} \ < \ 1 \, .
\label{caseiii}
\eea

\subsubsection*{2-body invariant mass variables} 
Since particle $A$ undergoes a 2-body decay, $m_{jj}^2$ is simply given by the resonance of particle $A$, i.e., $m_{jj}^2=m_C^2R_{AC}$. Due to the 3-body decay of $C$, the endpoint of $m_{\ell\ell}^2$ is given by
\bea
\left( m_{\ell\ell}^{\max} \right)^2 \ = \ m_C^2(1-\sqrt{R_{AC}})^2 \, .
\eea
For the purpose of deriving the analytic expressions of kinematic endpoints for $m_{\ell j}^>$ and $m_{\ell j}^<$, we first express the momenta for all visible particles in the rest frame of particle $C$:
\bea
p_{\ell_n}^{\mu}&\ = \ &\frac{m_{\ell\ell}}{2}(\cosh\eta_1+\sinh\eta_1 \cos\alpha,\;\sinh\eta_1+\cosh\eta_1 \cos\alpha,\; \sin\alpha,\;0) \, , \\
p_{\ell_f}^{\mu}&\ = \ &\frac{m_{\ell\ell}}{2}(\cosh\eta_1-\sinh\eta_1 \cos\alpha,\;\sinh\eta_1-\cosh\eta_1 \cos\alpha,\; -\sin\alpha,\;0) \, ,\\
p_j^{\mu}& \ = \ &\frac{m_A}{2}(\cosh\eta_2+\sinh\eta_2 \cos\theta,\;\sinh\eta_2+\cosh\eta_2 \cos\theta,\; \sin\theta \cos\phi,\;\sin\theta \sin\phi) \, ,
\eea
where $\eta_1$ and $\eta_2$ are defined as
\bea
\cosh\eta_1 \ \equiv \ \frac{m_C^2-m_A^2+m_{\ell\ell}^2}{2m_C m_{\ell\ell}}\, ,\;\;\; \cosh\eta_2 \ \equiv \  \frac{m_C^2+m_A^2-m_{\ell\ell}^2}{2m_C m_A}\, .
\eea
With these definitions, one can prove the following useful relations:
\bea
m_{\ell\ell} \sinh\eta_1 &\ = \ &-m_A \sinh\eta_2 \ = \ \frac{\lambda^{1/2}(m_C^2,\;m_A^2,\;m_{\ell\ell}^2)}{2m_C} \, ,
\eea
where $\lambda$ denotes the kinematic triangular function defined as
\bea
\lambda(x,\;y,\;z) \ \equiv \ x^2+y^2+z^2-2(xy+yz+zx) \, .
\eea
The symmetry of $\ell_n \leftrightarrow \ell_f$ enables us to have
\bea
x+z & \ = \ &\frac{\sqrt{y}}{2}\sqrt{R_{AC}}[\cosh(\eta_1-\eta_2)-\sinh(\eta_1-\eta_2)\cos\theta] \nonumber \\
& \ = \ &\frac{1}{2}\left(1-R_{AC}-y-\sqrt{(1+R_{AC}-y)^2-4R_{AC}}\cos\theta \right) \, ,
\eea
which is maximized at $\cos\theta=-1$ and $y=0$, i.e.,
\bea
(x+z)_{\max} \ = \ 1-R_{AC} \, .
\eea
From this constraint, we find that $m_{\ell j}^>$ can be maximized when either $x$ or $z$ vanishes, and $m_{\ell j}^<$ can be maximized when $x=z$. Therefore, we have
\bea
\left[ \left(m_{\ell j}^{>,\max}\right)^2,\; \left(m_{\ell j}^{<,\max}\right)^2\right] \ = \ m_C^2\left[1-R_{AC},\;\frac{1-R_{AC}}{2}\right] \, .
\eea

\subsubsection*{3-body invariant mass variables}
To obtain the kinematic endpoint for $m_{\ell\ell j}^2$  the relevant result in Ref.~\cite{KMP} can be reused:
\bea
\left(m_{\ell\ell j}^{\max}\right)^2 \ = \ m_C^2(1-R_{AC}) \, .
\eea
Again, for the other two variables, we can simply go through the same argument in Case (i) to obtain 
\bea
\left[ \left(m_{\ell jj}^{>,\max}\right)^2,\; \left(m_{\ell jj}^{<,\max}\right)^2\right] \ = \ m_C^2\left[1,\;\frac{1+R_{AC}}{2}\right] \, .
\eea
Note that the second expression corresponds to the kinematic configuration where the two visible particles are moving in the same direction with equal energy while particle $A$ is moving in the opposite direction.


\subsubsection*{Applying the formulae to Case (vii)} 
As in the previous two cases, we again reuse all expressions derived thus far with the simple replacement of $m_C^2$ by $s$. In this case, the only variable having no dependence on $s$ is $m_{jj}$ and all other variables unreliable as the chance of having $\hat{s}_{\max}=s$ is so small that the kinematic endpoints depending on $s$ are not typically saturated. With the assumption that $s$ is much larger than any of on-shell particle masses, we have $R_{AC} \rightarrow 0$. As $R_{BC}\geq 1$, $R_{AB}$ should be close to 0, giving the inequalities defining the corresponding region (denoted by $\mathcal{R}_7$) as follows:
\bea
R_{BC} \ \geq  \ 1 \quad \hbox{ and } \quad R_{AB} \ = \ 0 \, .
\label{casevii}
\eea
\subsection{Case (iv): $m_A, m_B \geq m_C$}

This scenario can be represented by a single 4-body decay of a heavy resonance $C$. A convenient frame for the relevant analysis is at the rest frame of particle $C$. The mass hierarchy of this case enables us to have the following inequality defining the corresponding region (denoted by $\mathcal{R}_4$):
\bea
R_{BC} \ \geq \ \max \left[\frac{1}{R_{AB}},\;1 \right] \quad \hbox{ and } \quad R_{AB}  >  0 \, .
\label{caseiv}
\eea

\subsubsection*{2-body invariant mass variables} 
Due to the 4-body decay feature of this process, $m_{jj}^2$ and $m_{\ell\ell}^2$ have an identical distribution whose kinematic endpoint is given by
\bea
\left(m_{jj}^{\max}\right)^2 \ = \ \left(m_{\ell\ell}^{\max}\right)^2 \ = \ m_C^2 \, .
\eea
For the maximum of $m_{\ell j}^>$, we imagine the situation where one of two $\ell$'s and one of two $j$'s are so soft in the rest frame of particle $C$ that all mass-energy of $C$ is split into the other $\ell$ and the other $j$. Similarly, for the maximum of $m_{\ell j}^<$, we imagine the situation where one of $j$ is extremely soft, the two leptons are emitted in the same direction, and the other $j$ is emitted in the opposite direction so that $m_{\ell_n j}^2=m_{\ell_f j}^2$ and $m_{\ell\ell}^2=0$. Therefore, we have
\bea
\left[ \left(m_{\ell j}^{>,\max}\right)^2,\; \left(m_{\ell j}^{<,\max}\right)^2\right] \ = \ m_C^2\left[1,\;\frac{1}{2}\right] \, .
\eea

\subsubsection*{3-body invariant mass variables}
The expression for the $m_{\ell\ell j}^2$ kinematic endpoint is simply given by
\bea
\left(m_{\ell\ell j}^{\max}\right)^2 \ = \ m_C^2 \, .
\eea
For the other two variables, we imagine the kinematic configuration where the two leptons are extremely soft while the two jets are emitted back-to-back in the rest frame of $C$.
\bea
\left(m_{\ell jj}^{>,\max}\right)^2 \ = \ \left(m_{\ell jj}^{<,\max}\right)^2 \ = \ m_C^2 \, .
\eea


\subsection{Summary}\label{sec:summary}
\begin{figure}[t]
\centering
\includegraphics[width=9cm]{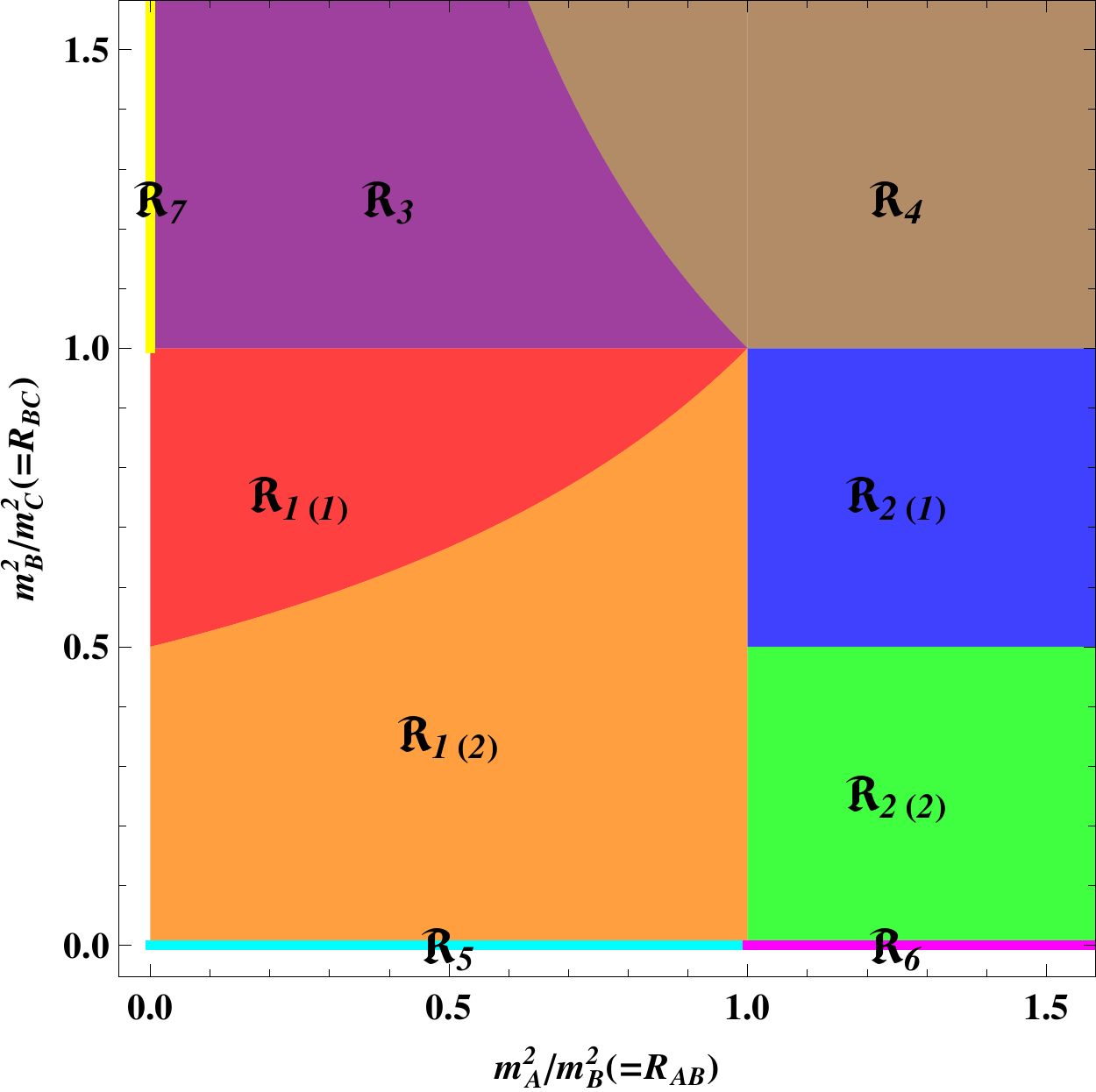}
\caption{\label{fig:regionplot} The parameter space division in $(R_{AB}, R_{BC})$ plane. The subscripted numbers correspond to various decay topologies in Figure~\ref{fig:decaytopo}, as listed in Table~\ref{tab:assignment}. Additional subscripts in the parenthesis denote subdivisions in the associated topology. }
\end{figure}

Let us collect the formulae we have derived so far for the kinematic endpoints and rearrange  them with respect to invariant mass variables in Eqs.~(\ref{eq:2body})--(\ref{eq:4body}). To make a connection of the formulae with specific LRSM scenarios, we again refer the readers to Table~\ref{tab:assignment}.
\bea
a \ \equiv \ \left( m_{\ell\ell}^{\max}\right)^2 & \ = \ & \left\{
\begin{array}{ll}
m_C^2(1-m_A^2/m_B^2)(1-m_B^2/m_C^2) & \hbox{ for }\mathcal{R}_{1(1)},\;\mathcal{R}_{1(2)} \\
s(1-m_A^2/m_B^2) & \hbox{ for }\mathcal{R}_5 \\
m_C^2(1-m_B^2/m_C^2) & \hbox{ for } \mathcal{R}_{2(1)},\; \mathcal{R}_{2(2)} \\
m_C^2(1-m_A/m_C)^2 & \hbox{ for }\mathcal{R}_3 \\
m_C^2 & \hbox{ for }\mathcal{R}_4\\
s & \hbox{ for } \mathcal{R}_6,\;\mathcal{R}_7 \\
\end{array}
\right. \label{eq:inva} \\
b \ \equiv \ \left( m_{\ell j}^{>,\max}\right)^2 & \ = \ & \left\{
\begin{array}{ll}
m_B^2(1-m_A^2/m_B^2) & \hbox{ for } \mathcal{R}_{1(1)} \\
m_C^2(1-m_B^2/m_C^2) & \hbox{ for } \mathcal{R}_{1(2)},\;\mathcal{R}_{2(2)} \\
m_B^2 & \hbox{ for }\mathcal{R}_{2(1)} \\
m_C^2 (1-m_A^2/m_C^2) & \hbox{ for } \mathcal{R}_3 \\
m_C^2 & \hbox{ for }\mathcal{R}_4 \\
s &\hbox{ for } \mathcal{R}_5,\;\mathcal{R}_6,\;\mathcal{R}_7
\end{array}
\right. \\
c \ \equiv \ \left(m_{\ell j}^{<,\max} \right)^2 & \ = \ & \left\{
\begin{array}{ll}
m_C^2(1-m_B^2/m_C^2) &\hbox{ for } \mathcal{R}_{1(1)},\;\mathcal{R}_{2(1)} \\
m_B^2(1-m_A^2/m_B^2) &\hbox{ for }\mathcal{R}_{1(2)},\;\mathcal{R}_5 \\
m_B^2 & \hbox{ for } \mathcal{R}_{2(2)},\mathcal{R}_6 \\
m_C^2(1-m_A^2/m_C^2)/2 & \hbox{ for } \mathcal{R}_3 \\
m_C^2/2  & \hbox{ for } \mathcal{R}_4 \\
s/2  & \hbox{ for } \mathcal{R}_7
\end{array}
\right. \\
d \ \equiv \ \left(m_{jj}^{\max}\right)^2& \ = \ & \left\{
\begin{array}{ll}
\left[m_A^2\right] & \hbox{ for }\mathcal{R}_{1(1)},\;\mathcal{R}_{1(2)},\;\mathcal{R}_3,\;\mathcal{R}_5,\mathcal{R}_7  \\
m_B^2       & \hbox{ for }\mathcal{R}_{2(1)},\;\mathcal{R}_{2(2)},\;\mathcal{R}_6 \\
m_C^2             & \hbox{ for }\mathcal{R}_4
\end{array}\right. \\
e \ \equiv \ \left(m_{\ell\ell j}^{\max}\right)^2 & \ = \ & \left\{
\begin{array}{ll}
m_C^2(1-m_A^2/m_C^2) & \hbox{ for } \mathcal{R}_{1(1)},\;\mathcal{R}_{1(2)},\mathcal{R}_3 \\
m_C^2 & \hbox{ for } \mathcal{R}_{2(1)},\;\mathcal{R}_{2(2)},\;\mathcal{R}_4 \\
s & \hbox{ for } \mathcal{R}_5,\;\mathcal{R}_6,\;\mathcal{R}_7
\end{array}
\right. \\
f \ \equiv \ \left(m_{\ell jj}^{>,\max}\right)^2 & \ = \ & \left\{
\begin{array}{ll}
m_B^2 & \hbox{ for }\mathcal{R}_{1(1)} \\
m_C^2(1-m_B^2/m_C^2+m_A^2/m_C^2) & \hbox{ for } \mathcal{R}_{1(2)} \\
m_C^2 & \hbox{ for } \mathcal{R}_{2(1)},\;\mathcal{R}_{2(2)},\mathcal{R}_3, \;\mathcal{R}_4\\
s & \hbox{ for } \mathcal{R}_5,\;\mathcal{R}_6,\;\mathcal{R}_7
\end{array}
\right. \\
g \ \equiv \ \left(m_{\ell jj}^{<,\max}\right)^2 & \ = \ & \left\{
\begin{array}{ll}
m_C^2(1-m_B^2/m_C^2+m_A^2/m_C^2) & \hbox{ for }\mathcal{R}_{1(1)} \\
m_B^2  & \hbox{ for } \mathcal{R}_{1(2)},\;\mathcal{R}_{2(1)},\;\mathcal{R}_{2(2)},\;\mathcal{R}_{5},\;\mathcal{R}_{6} \\
m_C^2(1+m_A^2/m_C^2)/2 & \hbox{ for } \mathcal{R}_3\\
m_C^2 & \hbox{ for }\mathcal{R}_4 \\
s/2 & \hbox{ for } \mathcal{R}_7
\end{array}
\right. \\
h \ \equiv \ \left(m_{\ell\ell jj}^{\max}\right)^2& \ = \ & \left\{
\begin{array}{ll}
\left[m_C^2 \right] & \hbox{ for }\mathcal{R}_{1(1)},\;\mathcal{R}_{1(2)},\;\mathcal{R}_{2(1)},\;\mathcal{R}_{2(2)},\;\mathcal{R}_{3},\;\mathcal{R}_{4} \\
s & \hbox{ for } \mathcal{R}_5,\;\mathcal{R}_6,\;\mathcal{R}_7
\end{array}\right. \label{eq:invh}
\eea
where the expressions in the square brackets indicate that the associated distributions appear as a resonance peak. Since the mass parameter $m_C^2$ merely determines the overall scale, the relevant parameter space can be divided in the $\left(\frac{m_A^2}{m_B^2}, \frac{m_B^2}{m_C^2}\right)=(R_{AB}, R_{BC})$ plane, as illustrated in Figure~\ref{fig:regionplot}. As introduced earlier, $\mathcal{R}_1$--$\mathcal{R}_7$ correspond to different decay topologies diagrammed in Figure~\ref{fig:decaytopo} and tabulated in Table~\ref{tab:assignment} [see also Eqs.~\eqref{casei}, \eqref{caseii}, \eqref{caseiii} and \eqref{caseiv}]. Additional subscripts in the parenthesis distinguish subregions in the associated decay topology  [cf.~Eqs.~\eqref{eq:caseiregion}, \eqref{mljj}, and \eqref{mlj2}]. Note that regions $\mathcal{R}_5$, $\mathcal{R}_6$, and $\mathcal{R}_7$ are represented by one-dimensional strips unlike the others. This is because the basic assumption that the center of mass energy $s$ is much larger than any of the masses of on-shell particles allows either $R_{AB}$ or $R_{BC}$ to vanish [cf.~Eqs~\eqref{casev}, \eqref{casevi}, and \eqref{casevii}].

\section{Topology disambiguation and mass measurement}\label{sec:topology}
In the busy environment of a real-life hadron collider experiment, leptons are much better-reconstructed than jets. Thus, from the precision measurement aspects, the best among the eight invariant mass variables listed in Eqs.~(\ref{eq:2body})--(\ref{eq:4body}) is $m_{\ell\ell}$. It therefore follows that as a minimal choice, the dilepton invariant mass variable can be used for topology disambiguation. For example, Ref.~\cite{Chen:2013fna} examined the shape of the dileption invariant mass distribution for the decay topologies \eqref{eq:LL}--\eqref{eq:RL}. This was based on the observation that different decay topologies involve different spin correlations~\cite{Han:2012vk} which affect the $m_{\ell\ell}$ spectrum. However, typical challenges in shape analysis are the facts that (relatively) large statistics is required and more importantly, the relevant shape would be severely distorted by the detailed dynamics and the predicament of hard cuts essential to suppress the relevant SM background.\footnote{Despite such potential difficulties, the invariant mass shape can in some cases be sufficient to completely determine the new particle mass spectrum, including the overall mass scale~\cite{Cho:2012er}.} On the contrary, the kinematic endpoint does not suffer from not only the details of the associated decay but the cuts to suppress backgrounds because it depends only on some specific kinematic configurations. Also, the need for ``local'' information near the endpoint typically demands less statistics than the shape analysis. In any case, we lose useful information encoded in the shape of the distribution, and the measurement of $m_{\ell\ell}$ endpoint by itself does not enable us to distinguish a certain topology from others. Hence, we need to supplement this with other observables.

\subsection{Constraints and Correlation matrix}
Following the basic idea of involving as few jets as possible in the observables to be used, the next best ones are the invariant mass variables having only a single jet, i.e., $m_{\ell j}^>$, $m_{\ell j}^<$, and $m_{\ell\ell j}$. However, including $m_{\ell\ell}$, this makes the number of observables greater than the number of unknown mass parameters ($m_C$, $m_B$, and $m_A$), namely, the system is now over-constrained.\footnote{For some specific cases of LRSM, either $A$ or $C$, or both can be the SM $W$ boson. One would then expect the number of unknown mass parameters reduced to two or one. However, since we do not know which underlying scenario governs the channel of our current interest, we generically treat the masses of $A,B,C$ as unknown parameters in our discussion. } We find several characteristic sum rules/constraints among the four invariant mass variables mentioned above, which are listed below in terms of the notations introduced in Eqs.~(\ref{eq:inva})--(\ref{eq:invh}):
\begin{itemize}
\item[C1:] $e=b+c$ for $\mathcal{R}_{1(1)}$, $\mathcal{R}_{1(2)}$, $\mathcal{R}_{2(1)}$, and $\mathcal{R}_{2(2)}$.
\item[C2:] $a=b$ for $\mathcal{R}_{2(2)}$ and $\mathcal{R}_{4}$.
\item[C3:] $a=c$ for $\mathcal{R}_{2(1)}$.
\item[C4:] $b=2c=e$ for $\mathcal{R}_{3}$.
\item[C5:] $a=b=2c=e$ for $\mathcal{R}_4$.
\item[C6:] Only $c$ is a well-defined endpoint for $\mathcal{R}_5$, $\mathcal{R}_6$, and $\mathcal{R}_7$.
\item[C7:] For $\mathcal{R}_{1(2)}$ and $\mathcal{R}_{5}$, $m_{\ell j}^>$ has a ``foot'' structure in it, i.e., a cusp is developed in the middle of the distribution. The cusp position should be the same as $c$.
\end{itemize}
We also provide a correlation matrix between regions and sum rules/constraints given above in Table~\ref{tab:corrmat}. Here the symbol ``$\surd$ '' implies that the relevant sum rule/constraint should be obeyed for the region of interest. We see that in principle, it is possible to discern all regions but $\mathcal{R}_{6}$ and $\mathcal{R}_{7}$ using C1--C7.
\begin{table}[t]
\centering
\begin{tabular}{c|c c c c c c c c c}
 & $\mathcal{R}_{1(1)}$ & $\mathcal{R}_{1(2)}$ & $\mathcal{R}_{2(1)}$ & $\mathcal{R}_{2(2)}$ & $\mathcal{R}_{3}$ & $\mathcal{R}_{4}$ & $\mathcal{R}_{5}$ & $\mathcal{R}_{6}$ & $\mathcal{R}_{7}$ \\
 \hline \hline
C1 & $\surd$ & $\surd$ & $\surd$ & $\surd$ & & & & & \\
C2 & & & & $\surd$ & & $\surd$ & & & \\
C3 & & & $\surd$ & & & & & & \\
C4 & & & & & $\surd$ & & & & \\
C5 & & & & & & $\surd$ & & & \\
C6 & & & & & & & $\surd$ & $\surd$ & $\surd$ \\
C7 & & $\surd$ & & & & & $\surd$ & & \\
\hline
C8 & $\surd$ & $\surd$ & & & $\surd$ & & $\surd$ & & $\surd$ \\
C9 & & & $\surd$ & $\surd$ & & $\surd$ & & $\surd$ &  \\
C10 & & & & & & $\surd$ & & & \\
C11 & $\surd$ & & & & & & & &
\end{tabular}
\caption{\label{tab:corrmat} Correlation matrix between regions and sum rules/constraints. ``$\surd$ '' implies that the relevant sum rule/constraint should be satisfied for the region of interest. C1--C7 involve only the invariant mass variables having 0 or 1 jet, while the others involve $\geq2$ jets.}
\end{table}

Of course, more sum rules/constraints can be utilized, once we allow the invariant mass variables involving more than 1 jet, which enable us to distinguish even $\mathcal{R}_{6}$ and $\mathcal{R}_{7}$. Instead of exhausting all possibilities, we enumerate some of them below:
\begin{itemize}
\item[C8:] $d$ appears as a resonance peak for $\mathcal{R}_{1(1)}$, $\mathcal{R}_{1(2)}$, $\mathcal{R}_{3}$, $\mathcal{R}_{5}$, and $\mathcal{R}_{7}$. 
\item[C9:] $d$ appears as a distribution for $\mathcal{R}_{2(1)}$, $\mathcal{R}_{2(2)}$, $\mathcal{R}_{4}$, $\mathcal{R}_{6}$.
\item[C10:] $a=d=f=g=h$ for $\mathcal{R}_4$.
\item[C11:] $af=bc$ for $\mathcal{R}_{1(1)}$.
\end{itemize}
It may be noted here that Refs.~\cite{Helo:2013dla,Helo:2013ika} studied a way of distinguishing event topologies (i) and (ii) in Figure~\ref{fig:decaytopo} by the existence of resonance peaks, which is relevant to C8 and C9.

\subsection{Mass measurement of new particles}

Once different regions are determined with the methods explicated in the previous section, the mass measurement of on-shell particles, or equivalently, the measurement of $R_{ij}$'s together with mass measurement of the first on-shell particle mass can be readily performed in terms of kinematic endpoints of various invariant mass distributions. We first provide the inverse formulae using only $a$, $b$, and $c$ for various regions but $\mathcal{R}_5$ and $\mathcal{R}_7$.
\bea
\mathcal{R}_{1(1)}:&& m_C^2 \ = \ \frac{c(a+b)}{a}\, ,\quad R_{BC} \ = \ \frac{b}{a+b}\, ,\quad  R_{AB} \ = \ \frac{c-a}{c}\, . \label{eq:inverse11} \\
\mathcal{R}_{1(2)}:&& m_C^2 \ = \ \frac{b(a+c)}{a}\, ,\quad R_{BC} \ = \ \frac{c}{a+c}\, ,\quad R_{AB} \ = \ \frac{b-a}{b}  \, .\\
\mathcal{R}_{2(1)}:&& m_C^2 \ = \ a+b\, ,\quad R_{BC} \ = \ \frac{b}{a+b} \, . \\
\mathcal{R}_{2(2)}:&& m_C^2 \ = \ a+c\, ,\quad R_{BC} \ = \ \frac{c}{a+c}\, . \\
\mathcal{R}_{3}:&& m_C^2 \ = \ \frac{(a+b)^2}{4a}\, ,\quad R_{AC} \ = \ \left(\frac{a-b}{a+b}\right)^2 \, . \\
\mathcal{R}_4:&& m_C^2 \ = \ a \, . \\
\mathcal{R}_6:&& m_B^2 \ = \ c \, .
\eea
For $\mathcal{R}_5$ and $\mathcal{R}_7$, the relevant mass parameters can be extracted by using other observables involving more than one jet. For example,
\bea
\mathcal{R}_5:&& m_B^2 \ = \ g\, ,\quad R_{AB} \ = \ \frac{g-c}{g} \, . \\
\mathcal{R}_7:&& m_A^2 \ = \ d \, . \label{eq:inverse7}
\eea
We should remark that Eqs.~(\ref{eq:inverse11})--(\ref{eq:inverse7}) are {\it not} unique ways of writing inverse formulae. For most of the regions, the interrelationships among the eight available kinematic endpoints are over-constrained due to less number of unknown mass parameters than the employed variables. Although the above-listed inverse formulae are expressed with a certain set of kinematic endpoints, others can as well be utilized for cross-checks.

\section{Application to the L-R seesaw}\label{sec:app}
In this section, we apply the general kinematic endpoint technique discussed in the preceding section to the specific case of L-R seesaw. Our aim is to illustrate the distinction between the invariant mass distributions for the scenarios mentioned in Eqs.~\eqref{eq:LL}--\eqref{eq:LR} and use the kinematic endpoints to disambiguate these scenarios from each other. First of all, we numerically verify the kinematic endpoint formulae derived in Section~\ref{sec:derivation} with phase-space decay event samples. The relevant distributions are illustrated in Figure~\ref{fig:theoryDist} of Appendix~\ref{sec:AppA}. Here we focus on a realistic collider simulation of the $\ell\ell jj$ signal in the context of various L-R seesaw model scenarios to demonstrate the effectiveness of the endpoint technique in probing the underlying event topology. We consider three benchmark scenarios (BS) to represent the parameter space points where 
$LL$, $RR$, and $R L$ 
modes are dominant respectively (see Section~\ref{sec:future} for details):
\bea
\hbox{BS1: } && m_{W_R}=5 \hbox{ TeV, }m_N=0.2 \hbox{ TeV, }|V_{\ell N}|^2=10^{-3} \, ,\\
\hbox{BS2: } && m_{W_R}=5 \hbox{ TeV, }m_N=4 \hbox{ TeV, }|V_{\ell N}|^2=10^{-6} \, ,\\
\hbox{BS3: } && m_{W_R}=5 \hbox{ TeV, }m_N=1 \hbox{ TeV, }|V_{\ell N}|^2=10^{-3} \, .
\eea
In terms of the model scenarios listed in Table~\ref{tab:assignment}, they correspond to $L_h L_h$, $R_lR_l$, and $R_l L_h$, respectively. One can therefore correspondingly relate them to event topologies (v), (ii), and (i), and hence, to regions $\mathcal{R}_5$, $\mathcal{R}_2$ and $\mathcal{R}_1$, respectively in Figure~\ref{fig:regionplot}. 

Parton-level events are generated with \texttt{MadGraph\_aMC@NLO}~\cite{Alwall:2014hca} and the parton distribution functions (PDFs) of protons are evaluated by the default \texttt{NNPDF2.3}~\cite{Ball:2012cx}. To describe parton showering and hadronization, the events are streamlined to \texttt{Pythia6.4}~\cite{Sjostrand:2006za}. The relevant output is subsequently fed into \texttt{Delphes3}~\cite{deFavereau:2013fsa} interfaced with \texttt{FastJet}~\cite{Cacciari:2011ma} for describing detector effects and finding jets. All the simulations are conducted for a $\sqrt{s}=14$ TeV $pp$ collider  at the leading order. In the first two benchmark scenarios, jets are formed using the anti-$k_t$ algorithm~\cite{Cacciari:2008gp} with a radius parameter $R=0.4$. For the last benchmark scenario, we point out that the on-shell $W$ bosons tend to be highly boosted due to a large mass gap between $N$ and $W$. The large boost of $W$ leads to a high collimation of the jets from its decay, thus requiring jet substructure techniques to resolve the subjets in each two-prong ``$W$''-jet. To this end, jets are initially clustered by the Cambridge-Aachen algorithm~\cite{Dokshitzer:1997in,Wobisch:1998wt} with a jet radius $R=1.2$, and the resulting ``fat'' jets are further processed by the Mass Drop Tagger method~\cite{Butterworth:2008iy}.


We emphasize that the main purpose of this simulation study is to see whether or not the proposed endpoint strategies are viable in the presence of detector effects. In this sense, including any potential backgrounds is beyond the scope of this paper, so we simply consider the signal component. Also, the precise measurement of kinematic endpoints typically requires large statistics; therefore, an arbitrarily large number of signal events are generated to minimize any statistical fluctuation. Finally, event selection is executed with a minimal set of selection criteria for simplicity. Of course, in the presence of backgrounds, one would impose a more stringent set of cuts to suppress them, but considering the fact that kinematic endpoints are typically least affected by cuts, we expect that the endpoint behaviors in our scheme will be similar to those under more sophisticated selection criteria.

Particle acceptance/isolation and detector geometry are basically performed according to the default parametrization in \texttt{Delphes3}~\cite{deFavereau:2013fsa}. We then choose the events having exactly two leptons (of either $e$ or $\mu$ flavor) and $\geq 2$ jets in the final state. For the second scenario (BS2), the first two hardest jets are used to construct relevant invariant masses, while for the other two scenarios (BS1 and BS3), the $W$ mass window is employed because the relevant jets were emitted from an on-shell $W$ boson. More specifically, we evaluate the dijet invariant mass for all possible combinations among observed jets and choose two jets satisfying a slightly tight condition of $|m_{jj}-m_W|<12 $ GeV. If there exist multiple combinations, then we simply take the one with the smallest difference. Although some events can contain more than 2 reconstructed jets, we evaluate the jet-involving invariant mass values using the {\it two} jets identified in the above-explained way, so that in every single event we have the same number of invariant mass values.
\begin{figure}[t]
\centering
\includegraphics[width=7cm]{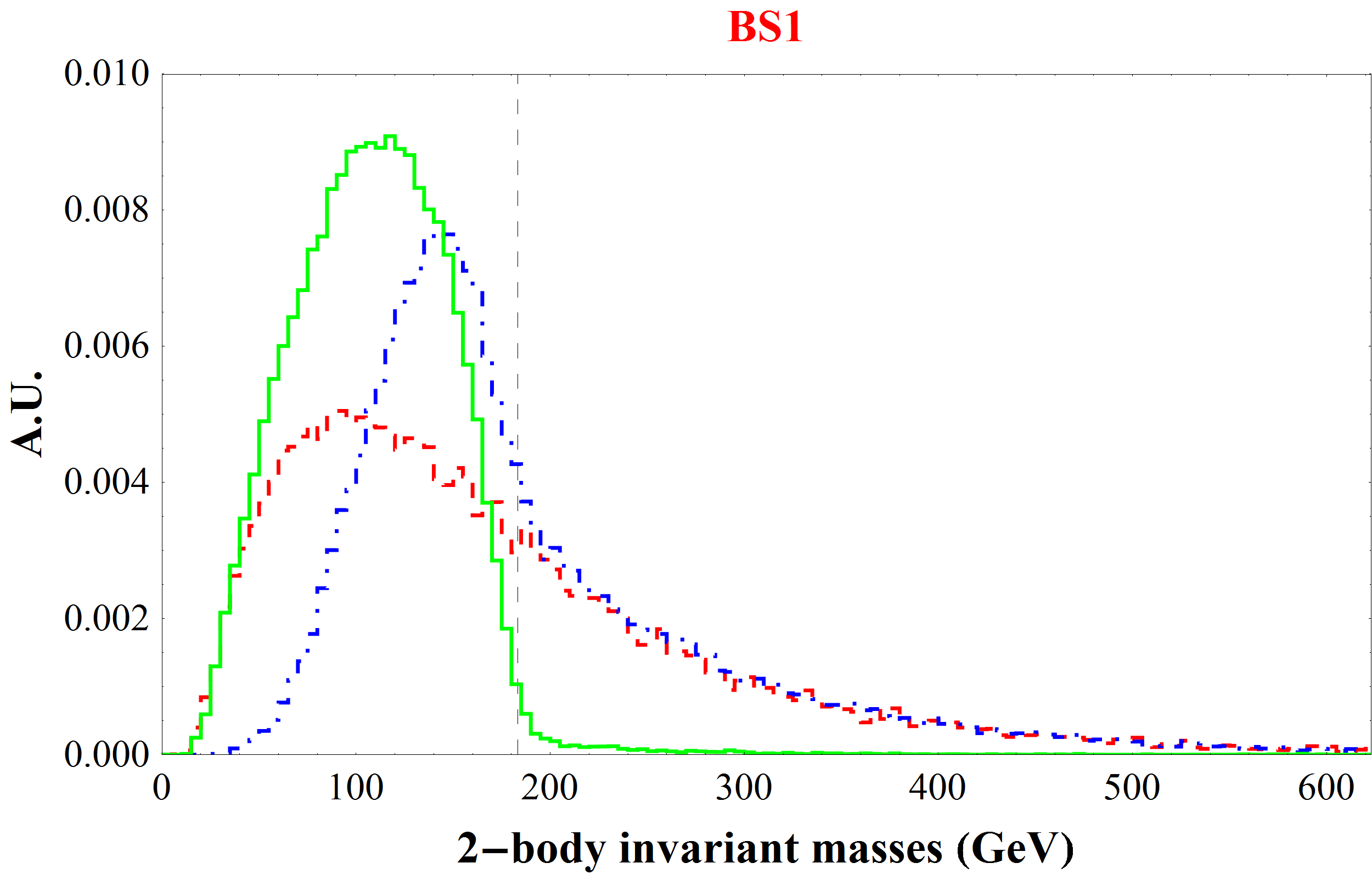}\hspace{0.5cm}
\includegraphics[width=7cm]{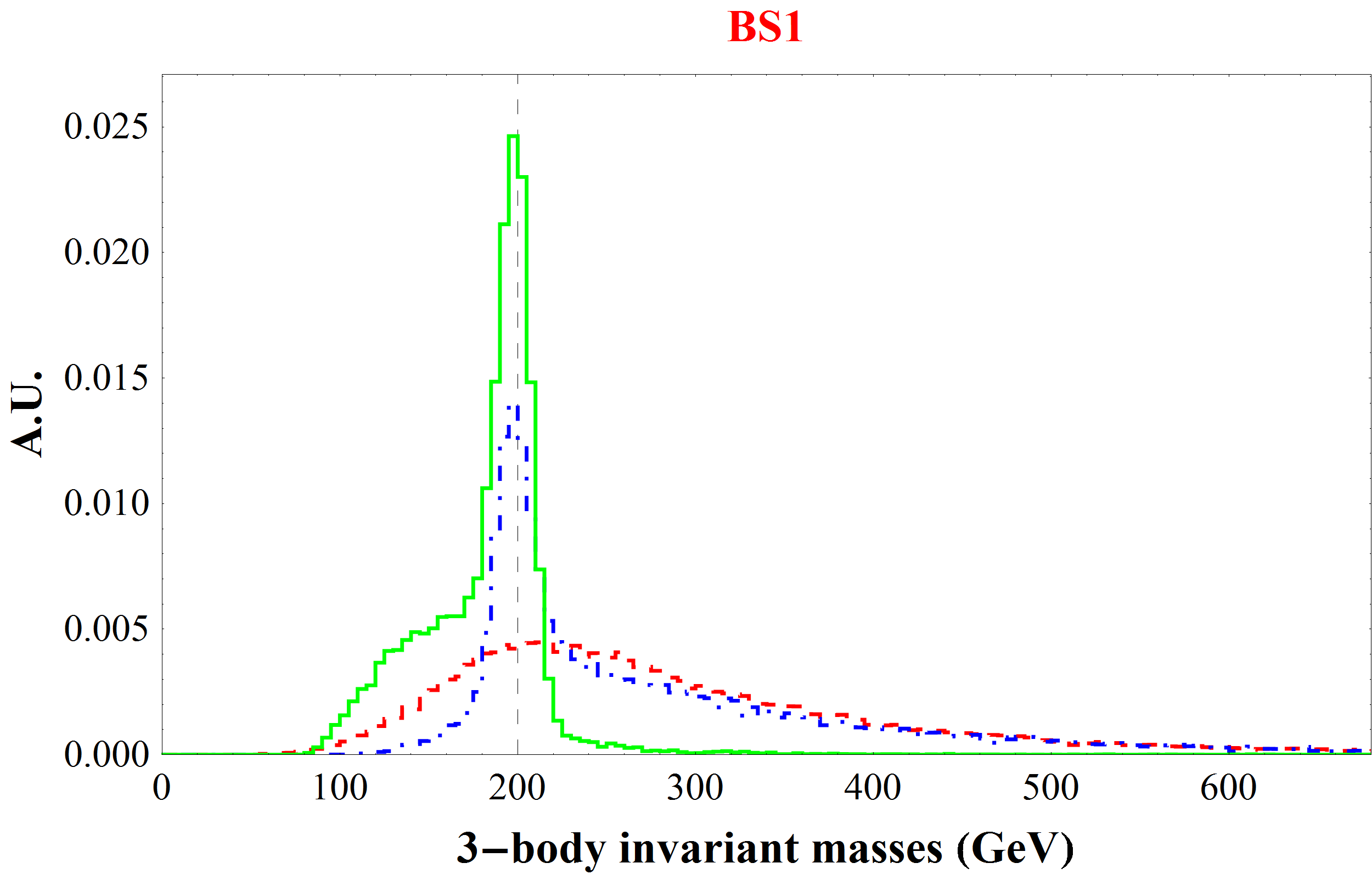}
\includegraphics[width=7cm]{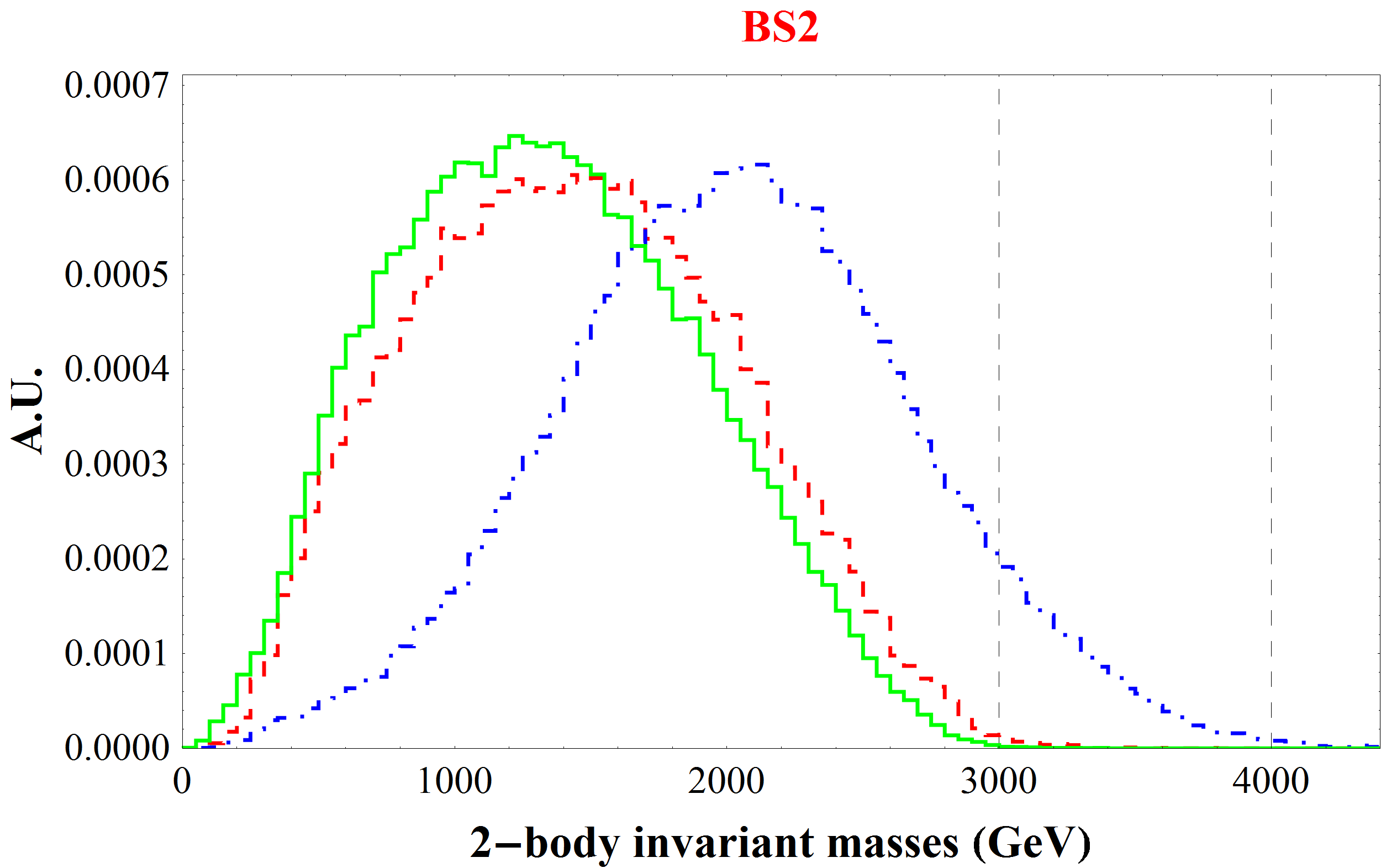}\hspace{0.5cm}
\includegraphics[width=7cm]{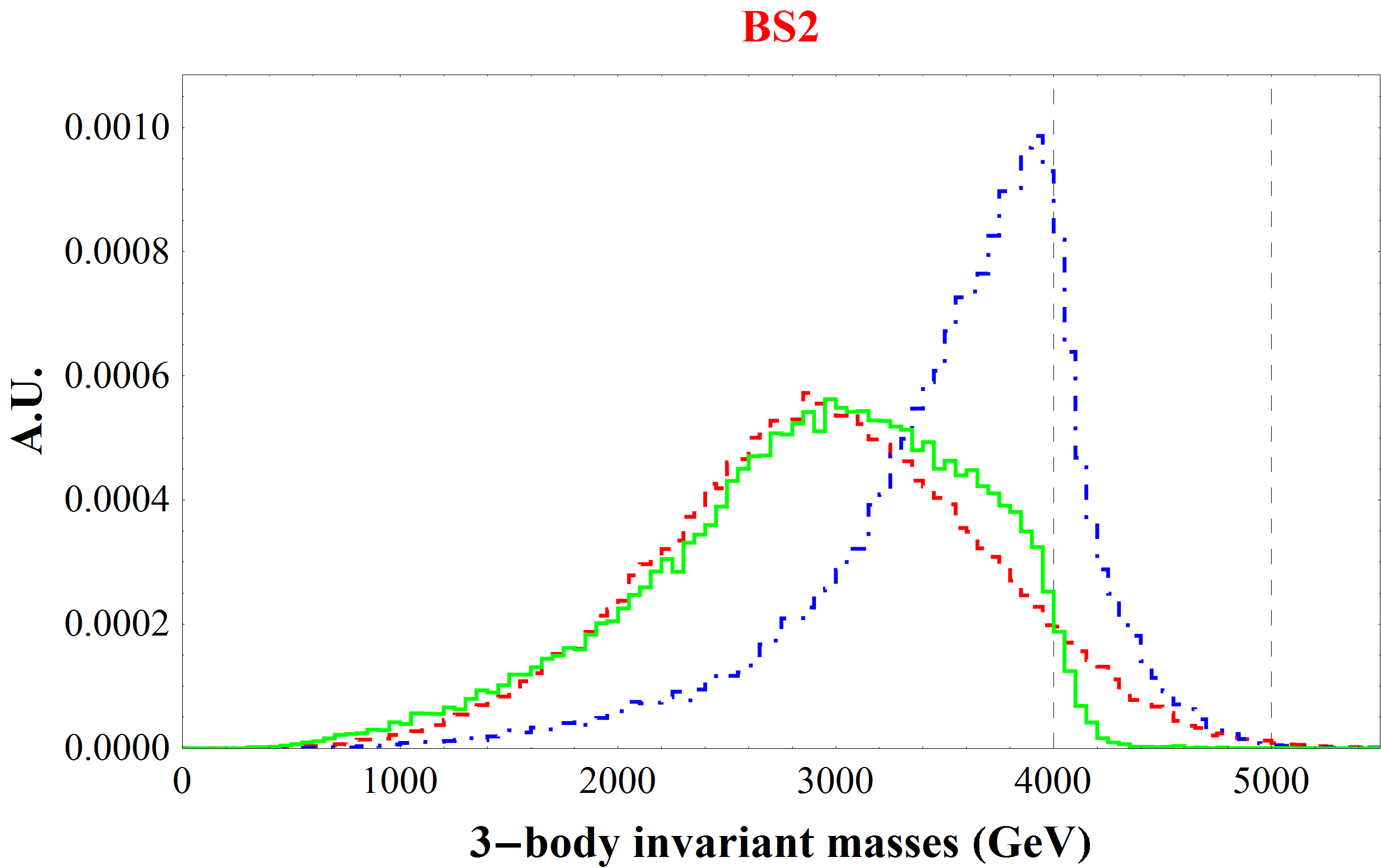}
\includegraphics[width=7cm]{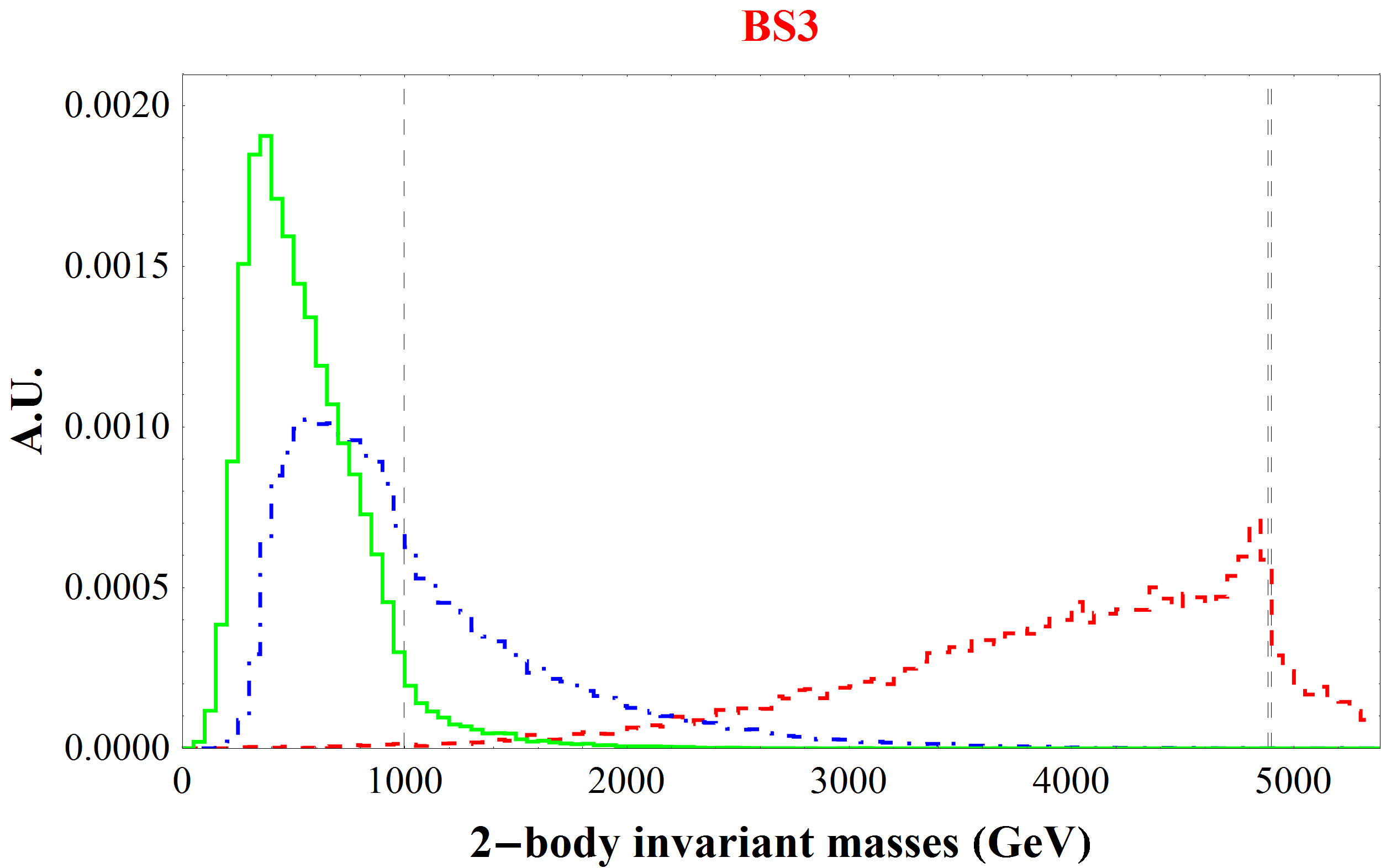}\hspace{0.5cm}
\includegraphics[width=7cm]{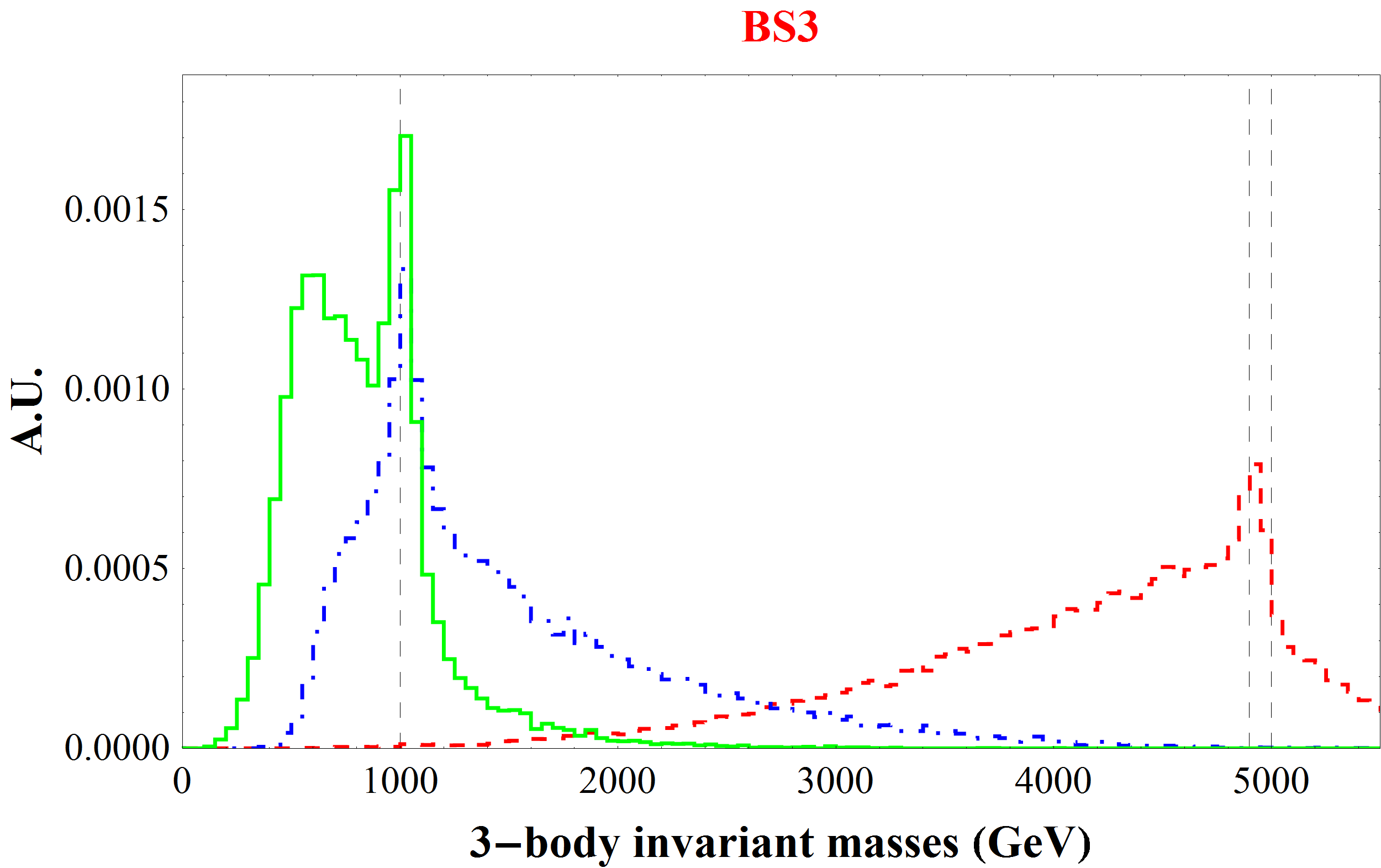}
\caption{\label{fig:detectorResults} Various invariant mass distributions for BS1 (top panels), BS2 (middle panels), and BS3 (bottom panels). The left panels display 2-body invariant mass variables such as $m_{\ell\ell}$ (red dashed), $m_{\ell j}^>$ (blue dot-dashed), and $m_{\ell j}^<$ (green solid),  while the right ones display 3-body invariant mass variables such as $m_{\ell\ell j}$ (red dashed), $m_{\ell jj}^>$ (blue dot-dashed), and $m_{\ell jj}^<$ (green solid). Black dashed lines denote the relevant theoretical kinematic endpoints as formulated in Section~\ref{sec:derivation}. The $y$-axis has an arbitrary unit. }
\end{figure}

We show various 2- and 3-body invariant mass distributions for the three benchmark scenarios in Figure~\ref{fig:detectorResults}.\footnote{Dijet invariant mass distributions are not displayed in the figure, as it develops a trivial distribution, i.e., a resonance peak, for BS1 and BS3.} Top, middle, and bottom panels correspond to BS1, BS2, and BS3, respectively. In each row of panels, the left panel shows the 2-body invariant mass distributions $m_{\ell\ell}$ (red dashed), $m_{\ell j}^>$ (blue dot-dashed), and $m_{\ell j}^<$ (green solid) while the right panel shows the 3-body invariant mass distributions $m_{\ell\ell j}$ (red dashed), $m_{\ell jj}^>$ (blue dot-dashed), and $m_{\ell jj}^<$ (green solid). Black dashed lines indicate the relevant theory predictions for the kinematic endpoints derived in Section~\ref{sec:derivation}. Note that we provide unit-normalized distributions for illustration since our focus is the structure of kinematic endpoints. Based on the analysis scheme described above, one can easily infer the corresponding relative weights. We observe that for BS1 and BS2 the invariant mass variables involving $\leq 1$ jet, i.e., $m_{\ell\ell}$, $m_{\ell j}^>$, $m_{\ell j}^<$, and $m_{\ell\ell j}$, are reasonably well-matched to the associated theoretical predictions in spite of detector effects such as jet energy resolution, smearing and contamination from initial and final state radiations (ISR/FSR).\footnote{Note that for BS1 the kinematic endpoints for $m_{\ell\ell}$, $m_{\ell j}^>$, and $m_{\ell\ell j}$ are ill-defined because their values can be arbitrarily large up to the associated kinematic limit determined by the available center of mass energy.} In contrast, once more jets are involved, i.e., $m_{\ell jj}^>$ and $m_{\ell jj}^<$, kinematic endpoints are more smeared so that identifying their correct positions would be rather challenging. When it comes to BS3, the situation becomes more challenging, and even for 2-body invariant mass distributions, the endpoints either suffer from relatively large smearing (green and red histograms) or are unsaturated (blue histogram). In particular, the information of subjets in the $W$-jet is typically less accurate than that of regular jets, thus rendering the distributions more smeared. Certainly, these phenomena can be improved by better jet energy measurement and ISR/FSR identification (see, for example, Ref.~\cite{Kim:2014ana} studying the impact of various detector effects upon the distributions of mass variables). Moreover, the boosted jet technique is a research field that is being actively investigated and developed. We therefore expect that locating the true kinematic endpoints will be advanced in the future, which will help us in determining the new particle masses more accurately. 

\section{L-R seesaw phase diagram}\label{sec:future}
In this section, we present some future prospects of probing the L-R seesaw parameter space using the different collider signals mentioned earlier. To be concrete, we only focus on the mass hierarchy $m_{W_R} > m_N>m_W$ which can be effectively probed at the LHC via the $\ell\ell jj$ final state.  In this case, as mentioned in Eqs.~\eqref{eq:LL}-\eqref{eq:LR}, there are four classes of Feynman diagrams for the $\ell\ell jj$ signal (Figure~\ref{fig1})~\cite{Chen:2013fna}. 
The $LL$ channel is a clear probe of the seesaw matrix in both SM seesaw 
and L-R seesaw models. However, its effectiveness solely relies on the heavy-light neutrino mixing parameter $|V_{\ell N}|^2$, and is limited to heavy neutrino mass $M_N$ only up to a 
few hundred GeV~\cite{Datta:1993nm,Dev:2013wba, Alva:2014gxa, Panella:2001wq, Han:2006ip, Aguila:2007em, Aguila:2008cj}. Experimentally, the mass range $M_N=100$--500 GeV has been explored at the $\sqrt s=8$ TeV LHC for $\ell=e,\mu$~\cite{Khachatryan:2015gha, Aad:2015xaa}, and direct upper limits on $|V_{\ell N}|^2$ of the order of $10^{-2}$--$10^{-1}$ have been set. 
We note here that the current indirect limits from a global fit of electroweak precision data (EWPD), lepton flavor violation, and lepton universality constraints~\cite{Antusch:2014woa} are roughly one to two 
orders of magnitude stronger.  For a bird's-eye view of other complementary  limits and the future sensitivities, see e.g.,~\cite{Deppisch:2015qwa}.  

In the $RR$ channel~\cite{Keung:1983uu}, the RH neutrino is produced on-shell due to its gauge interaction with $W_R$ [cf.~Eq.~\eqref{lry}] and subsequently decays into a 3-body final state via an off-shell $W_R$. This diagram only relies on the gauge coupling $g_R$ and is independent of $V_{\ell N}$; therefore, it gives the dominant contribution for small $V_{\ell N}$ which is of course the naive 
expectation in the ``vanilla'' type-I seesaw case. Using this 
channel, LHC exclusion limits are derived in the ($m_N,m_{W_R}$) 
plane, and currently exclude $m_{W_R}$ up to 3 TeV assuming the equality between the $SU(2)_R$ and $SU(2)_L$ gauge couplings, i.e., $g_R=g_L$~\cite{Aad:2015xaa, Khachatryan:2014dka}.\footnote{Similar lower limits on $m_{W_R}$ are also obtained from low-energy flavor changing neutral current observables~\cite{Bertolini:2014sua}. } In general, the signal cross section will be scaled by a factor of $(g_R/g_L)^4$, and hence, the corresponding limit on $m_{W_R}$ could be weaker if $g_R<g_L$. In any case, being independent of the Dirac 
neutrino Yukawa coupling, the $RR$ process 
does {\em not} probe the complete seesaw matrix. 

The $RL$ and $LR$ contributions, on the other hand, necessarily involve the heavy-light neutrino mixing. In fact, the $RL$ diagram could give the dominant contribution to the $\ell \ell jj$ signal, if the mixing $|V_{\ell N}|^2$ is not negligible 
and/or the $W_R$ gauge boson is not too heavy~\cite{Chen:2013fna}. There are two reasons for this 
dominance: (i) it leads to a production rate $\sigma(pp\to W_R\to N\ell^\pm)$ which is independent of mixing  and only suppressed by 
$(g_R/g_L)^4(m_{W}/m_{W_R})^4$ (as in the $RR$ case), and can therefore dominate over the 
$LL$ contribution which depends on $|V_{\ell N}|^2 g_L^4$; (ii) the RH 
neutrino in this case has a 2-body decay: $N\to \ell^\pm W_L \to \ell^\pm jj$ (as in the $LL$ case) which is not suppressed by the phase space, unlike in the $RR$ case with a 3-body decay of $N$. Hence, for a sizable range of the mixing and RH gauge boson mass, 
the $RL$ mode is expected to be dominant for the $\ell \ell jj$ signal at the LHC and could constitute a clear probe of the full seesaw matrix. There exist classes of L-R seesaw models where such large mixing parameters can be realized without much fine-tuning~\cite{Dev:2013oxa}.

The remaining possibility, namely, the $LR$ contribution is doubly suppressed by the heavy-light mixing as well as phase space, and hence, always 
smaller than one of the other three contributions discussed above. So this is not relevant for the experimental searches of L-R seesaw.  

\begin{figure}[t!]
\centering
\includegraphics[width=8cm]{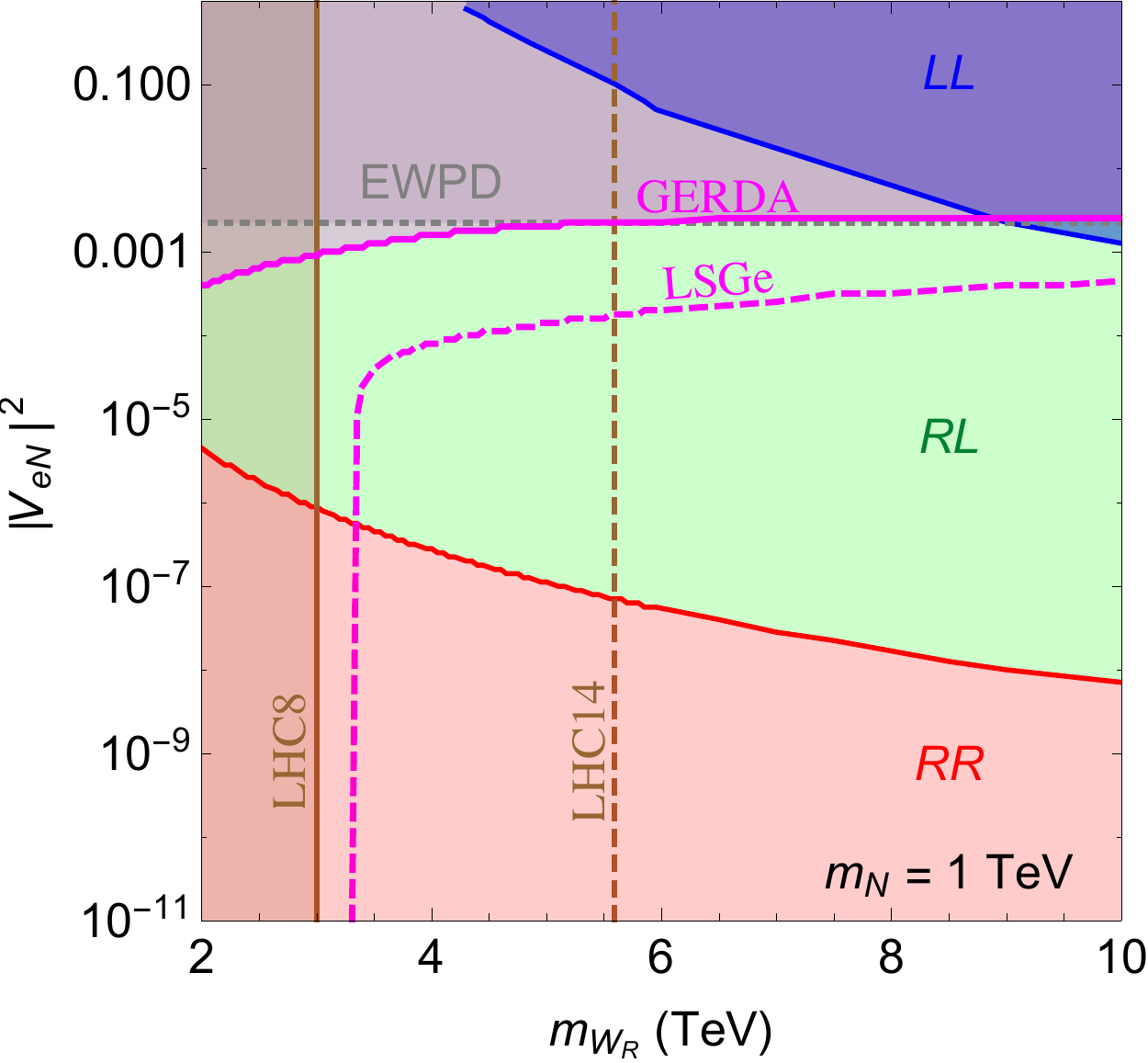}
\caption{L-R seesaw phase diagram for the $\ell \ell jj$ signal at the LHC with $m_N=1$ TeV. The regions denoted by $LL$ (blue), $RL$ (green) and $RR$ (red) correspond to the dominance of the signal from the respective channel. The solid (dashed) vertical line shows the current (future) lower limit on $m_{W_R}$ from the LHC, whereas the horizontal (grey) dotted line shows the indirect limit from EWPD. The magenta solid (dashed) curve shows the current (future) limit from $0\nu\beta\beta$.}
\label{fig:phase1}
\end{figure}

The regions of dominance for the different contributions discussed above are 
shown in Figure~\ref{fig:phase1} by various shaded regions (blue for $LL$, green for $RL$ and red for $RR$)  as a function of the mixing parameter in the electron sector for a fixed value of the heavy neutrino mass $m_N=1$ TeV and assuming $g_R=g_L$.  The vertical (brown) solid line in Figure~\ref{fig:phase1} shows the 95\% CL direct  limit on $m_{W_R}$ from the $\sqrt s=8$ TeV LHC~\cite{Khachatryan:2015gha}, whereas the vertical dashed line shows a projected lower limit from $\sqrt s=14$ TeV LHC with $300~{\rm fb}^{-1}$ integrated luminosity~\cite{Ferrari:2000sp}. For comparison, we also show the current 90\% CL upper limit on the mixing parameter from a recent global fit to the EWPD~\cite{Antusch:2014woa} (horizontal dotted line). All the signal cross sections for $LL$, $RL$ and $RR$ used in Figure~\ref{fig:phase1} are generated at the parton level for $\sqrt s=14$ TeV LHC again using  \texttt{MadGraph\_aMC@NLO}~\cite{Alwall:2014hca} with the default \texttt{NNPDF2.3}~\cite{Ball:2012cx} PDF sets. We have applied some basic common selection cuts for all the signals:
\begin{align}
p_T^{\ell}, ~p_T^j > 30~{\rm GeV}, \quad \slashed{E}_T <40~{\rm GeV}, \quad |\eta^j|<2.8,\quad |\eta^\ell| <2.5,\quad \Delta R^{jj},\Delta R^{j\ell} >0.4.
\end{align}
For simplicity, we assume all the non-standard Higgs bosons in the LRSM to be heavier than $W_R$, so that the total width of $W_R$ is mostly governed by the masses of $W_R$ and $N$.

The presence of Majorana neutrinos in the LRSM also gives several additional contributions to the low-energy LNV process of $0\nu\beta\beta$~\cite{Mohapatra:1979ia, Mohapatra:1981pm, Picciotto:1982qe, Hirsch:1996qw, Tello:2010am, Chakrabortty:2012mh, Nemevsek:2012iq, Barry:2013xxa, Dev:2013vxa, Dev:2013oxa, Huang:2013kma, Dev:2014xea, Borah:2015ufa, Ge:2015yqa, Awasthi:2015ota}. In the limit of large $m_{W_R}$, the only dominant contributions are due to LH current exchange, which gives an upper limit on the mixing parameter, independent of $m_{W_R}$, as shown by the solid magenta line in Figure~\ref{fig:phase1}. On the other hand, for very light RH gauge bosons, the purely RH current exchange diagram becomes dominant over the rest and puts a lower bound on $m_{W_R}$ for a given $m_N$ from the non-observation of $0\nu\beta\beta$, irrespective of the value of the mixing parameter, as shown by the vertical portion of the solid magenta curve. Here we have used the $^{76}$Ge isotope and the corresponding combined 90\% CL limit on the $0\nu\beta\beta$ half-life from GERDA phase-I: $T_{1/2}^{0\nu}> 3\times 10^{25}$ yr~\cite{Agostini:2013mzu} for illustration. For the nuclear matrix elements, we use the maximum values from a recent SRQRPA calculation~\cite{Meroni:2012qf}, so as to obtain the minimum half-life predictions. The dashed magenta curve shows the future sensitivity of multi-ton scale $0\nu\beta\beta$ experiments (LSGe) with $^{76}$Ge isotope, such as the proposed Majorana+GERDA experiment with an ultimate limit of $T_{1/2}^{0\nu}> 10^{28}$ yr~\cite{Abgrall:2013rze}.

\begin{figure}[t!]
\centering
\includegraphics[width=10cm]{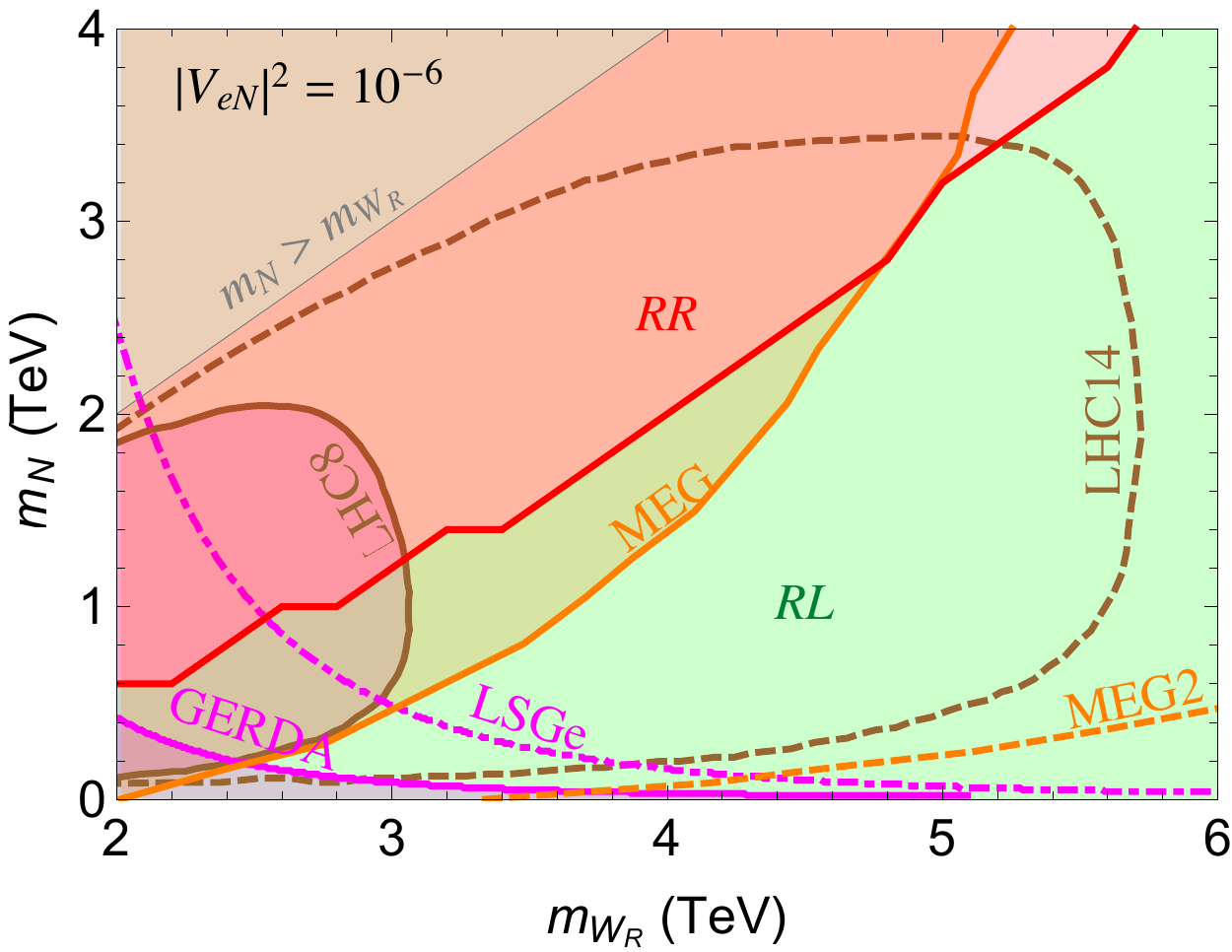}
\caption{L-R seesaw phase diagram for the $\ell \ell jj$ signal at the LHC with $|V_{eN}|^2=10^{-6}$. The labels are the same as in Figure~\ref{fig:phase1}. In addition, we have shown the exclusion region due to LFV constraints from MEG (solid orange) and the future limit from MEG2 (dashed orange). The grey shaded region (upper left corner) corresponds to $m_N>m_{W_R}$ which cannot be probed by the $\ell \ell jj$  signal at the LHC, but accessible to the low-energy searches.}
\label{fig:phase14}
\end{figure}
The phase diagram shown in Figure~\ref{fig:phase1} can be easily translated to the more familiar $(m_{W_R},m_N)$ parameter space, as shown in Figure~\ref{fig:phase14}.  The current 95\% CL exclusion contour from $\sqrt s=8$ TeV LHC~\cite{Khachatryan:2015gha} (LHC8) and the future sensitivity of $\sqrt s=14$ TeV LHC with $300~{\rm fb}^{-1}$ luminosity (LHC14)~\cite{Ferrari:2000sp} are also shown. Here we have fixed the mixing parameter $|V_{eN}|^2=10^{-6}$ for illustration. Due to the relatively small value of mixing, the $LL$ channel is no longer relevant, and either $RL$ or $RR$ channel is the dominant one in the entire L-R seesaw parameter space of interest. In particular, for smaller $m_N$ values,  the $RR$ channel becomes less efficient, compared to the $RL$ channel, due to the phase space suppression factor of $m_N^5/m_{W_R}^4$ in the 3-body decay rate $\Gamma(N\to \ell jj)$. Increasing the value of the mixing parameter will further enlarge the $RL$ dominance region. Thus, a combination of $RR$ and $RL$ modes provides a better probe of the L-R seesaw, as compared to the $RR$ mode alone. It is worth noting here that the $0\nu\beta\beta$ searches provide a complementary probe of the L-R seesaw parameter space, especially for the $m_N>m_{W_R}$ regime, which is kinematically inaccessible in the $\ell\ell jj$ channel at the LHC. We should note here that the $0\nu\beta\beta$ analogs of the $LL$, $RL$ and $RR$ modes will give rise to different angular distributions, which can in principle be measured in the proposed SuperNEMO experiment~\cite{Arnold:2010tu}. Similarly, one can use polarized beams in a linear collider to distinguish between the different contributions in L-R seesaw~\cite{Barry:2012ga}. These are complementary to the kinematic endpoint method proposed here for a hadron collider.

Another complementary low-energy probe of the LRSM is through the LFV processes~\cite{Cirigliano:2004mv, Cirigliano:2004tc, Das:2012ii, Barry:2013xxa, Awasthi:2015ota}.  In particular, the $\mu\to e\gamma$ decay rate gets an additional contribution from the purely RH current and is currently constrained to be ${\rm Br}(\mu\to e\gamma)<5.7\times 10^{-13}$ at 90\% CL by the MEG experiment~\cite{Adam:2013mnn}. Assuming a maximal mixing between the RH electron and muon sectors of the heavy neutrino mass matrix, we translate this limit into an exclusion region in the $(m_{W_R},m_N)$ parameter space, as shown by the shaded region above the solid orange curve in Figure~\ref{fig:phase14}. The projected  limit of ${\rm Br}(\mu\to e\gamma)<10^{-14}$ from the future upgrade of MEG experiment~\cite{Baldini:2013ke} could probe most of the remaining parameter space shown in Figure~\ref{fig:phase14}. This clearly illustrates the importance of a synergistic approach at both energy and intensity frontiers in testing the L-R seesaw paradigm in future.

\begin{figure}[t!]
\centering
\includegraphics[width=10cm]{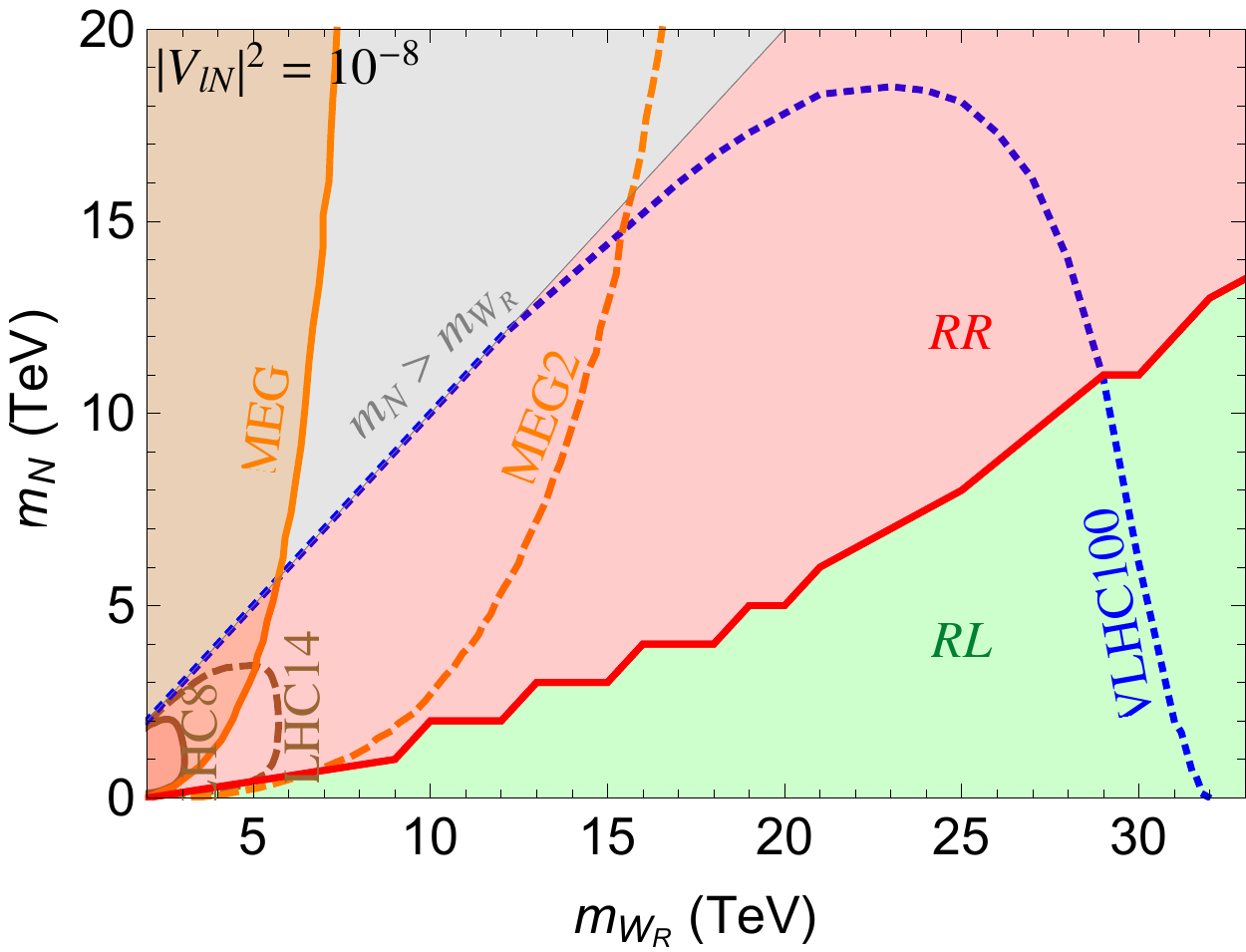}
\caption{Sensitivity reach of a futuristic $\sqrt s=100$ TeV collider for probing the L-R seesaw parameter space through the $\ell \ell jj$  signal. Other labels are the same as in Figure~\ref{fig:phase14}.}
\label{fig:phase100}
\end{figure}

Since the high energy physics community has seriously started thinking about a next-generation $\sqrt s=100$ TeV hadron collider~\cite{Hinchliffe:2015qma}, we feel motivated to present the sensitivity reach of such a machine in the context of L-R seesaw. As before, we generate the signal cross sections at the parton level using  \texttt{MadGraph\_aMC@NLO}~\cite{Alwall:2014hca} with the default \texttt{NNPDF2.3}~\cite{Ball:2012cx} PDF sets. We have used the following conservative selection cuts in our event simulation: 
\begin{align}
& p_T^{\ell,~{\rm leading}}\ > \ 60~{\rm GeV}\, , \quad p_T^{\ell,{\rm trailing}},~p_T^j \ > \ 40~{\rm GeV}, \quad \slashed{E}_T \ < \ 40~{\rm GeV}, \nonumber \\
& |\eta^j| \ < \ 2.8,\quad |\eta^{\ell}| \ < \ 2.5,\quad \Delta R^{jj},\Delta R^{j\ell} \ > \ 0.4 \, .
\end{align}
 Our results for the projected $2\sigma$ exclusion region with $1~{\rm ab}^{-1}$ integrated luminosity at $\sqrt s=100$ TeV are shown in Figure~\ref{fig:phase100} (blue dotted curve), which suggests that one could probe RH gauge boson masses up to 32 TeV, in good agreement with previous studies~\cite{Rizzo:2014xma}.  This sensitivity reach extends to regions of the LRSM parameter space not accessible by even future low-energy searches at the intensity frontier, as demonstrated by a  comparison with the MEG-2 projected limit (orange dashed curve) in Figure~\ref{fig:phase100}. Moreover, one can access even smaller mixing parameters at the collider through the $RL$ mode, as illustrated here for $|V_{\ell N}|^2=10^{-8}$ which still gives a dominant $RL$ contribution for relatively smaller $m_N$ values. Note that the shape of our exclusion contour for the $\sqrt s= 100$ TeV case is not exactly similar to the $\sqrt s=8$ or 14 TeV contours in the low $m_N$ region, mainly because we have not applied any specialized selection cuts on the invariant masses and have taken the tagging efficiencies to be 100\%, which is usually not the case in a realistic hadron collider environment. Nevertheless, our parton-level estimates should serve as a rough guideline for more sophisticated studies in future.    

\section{Conclusion \label{sec:conclusions}} 
In summary, we have pointed out a clean and robust way to distinguish between different mechanisms for the  production of dilepton plus dijet final states at a hadron collider using the kinematic endpoints of various invariant mass distributions. 
We derived analytic expressions for these kinematic endpoints, under a minimal set of assumptions : (i) no invisible particles are involved in the relevant process (i.e., no missing transverse momentum at the parton level), (ii) all the final state particles originate from a common mother particle, and (iii) the two jets are the decay products of the same particle. We then provided various criteria to distinguish the possible scenarios yielding the same $\ell\ell jj$ collider signature from one another. As a ``spin-off'', the potential determination of the masses of the heavy resonances in the associated process was also discussed. We emphasize that the relevant derivations, prescriptions and strategy are rather general and can be straightforwardly extended to other event topologies, even with invisible particles (see e.g., Ref.~\cite{KMP}).  

As a proof of principle, we have applied this general method to study the distinction between the seesaw models based on the Standard Model and the Left-Right symmetric model. Once there is statistically significant evidence for such a $\ell\ell jj$ signal, we can pinpoint the diagram(s) responsible for this signal by using our kinematic endpoint method and also measure the masses of the new particles involved in this process. This can be a powerful way to study the origin of neutrino masses at the LHC and beyond. Along this line, we examined the signal sensitivity for some well-motivated channels at the LHC and a 100 TeV future collider, and found that a $W_R$ gauge boson mass up to $\sim 5.5$ TeV with 300 fb$^{-1}$ data at $\sqrt s=14$ TeV LHC or up to $\sim 32$ TeV with 1 ab$^{-1}$ data at $\sqrt s=100$ TeV collider.  This has important consequences for other new physics observables, such as neutrinoless double beta decay, lepton flavor violation and even matter-antimatter asymmetry~\cite{Frere:2008ct, Dev:2014iva}.  

\section*{Acknowledgements}

We would like to thank Chengcheng Han, Konstantin Matchev and Myeonghun Park for useful discussions. D.K. would also like to thank the Center for Theoretical Underground Physics and Related Areas (CETUP*) for
its hospitality during the completion of this work. P.S.B.D. acknowledges the local hospitality and partial support of the Institute for Research in Fundamental Sciences (IPM), Tehran, where part of this work was done. This work of P.S.B.D. was supported in part by a TUM University Foundation Fellowship and the DFG cluster of excellence ``Origin and Structure of the Universe". The work of D.K. is supported in part by DOE Grant No. DE-SC0010296. The work of R.N.M. is supported in part by the National Science Foundation Grant No. PHY-1315155. D.K. also acknowledges the support by the LHC Theory Initiative postdoctoral fellowship (NSF Grant No.~PHY-0969510). 

\appendix

\section{Invariant mass distributions \label{sec:AppA}}
\begin{figure}[t]
\centering
\includegraphics[width=3.6cm]{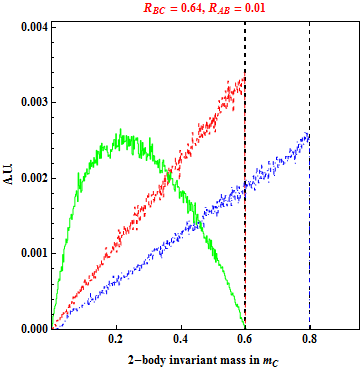}
\includegraphics[width=3.6cm]{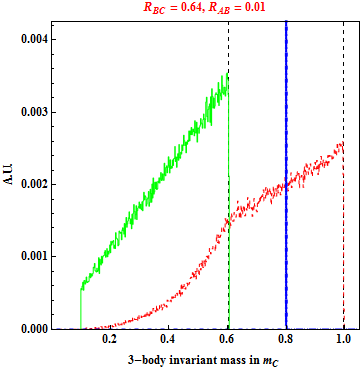}
\includegraphics[width=3.6cm]{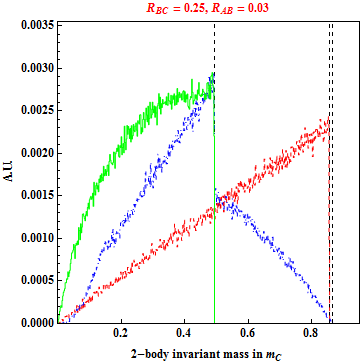}
\includegraphics[width=3.6cm]{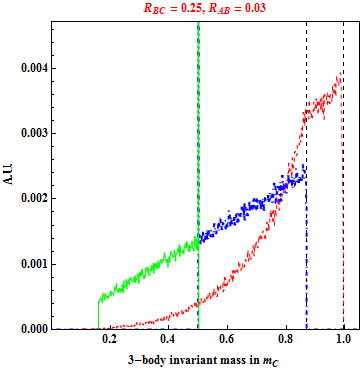} \\
\includegraphics[width=3.6cm]{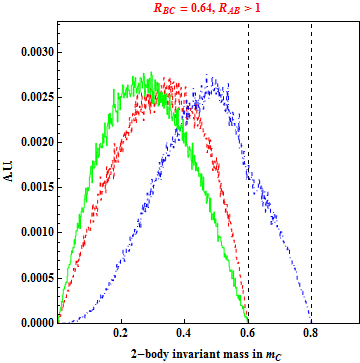}
\includegraphics[width=3.6cm]{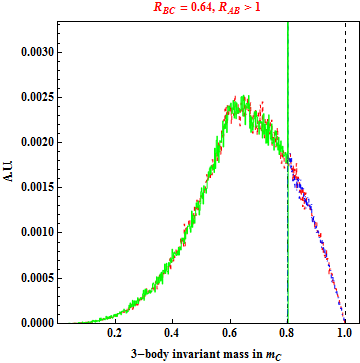}
\includegraphics[width=3.6cm]{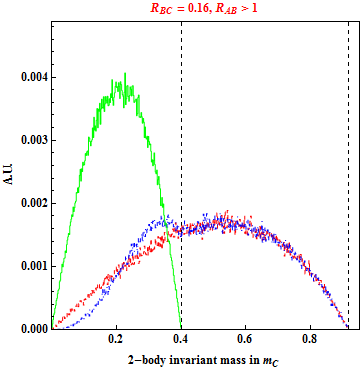}
\includegraphics[width=3.6cm]{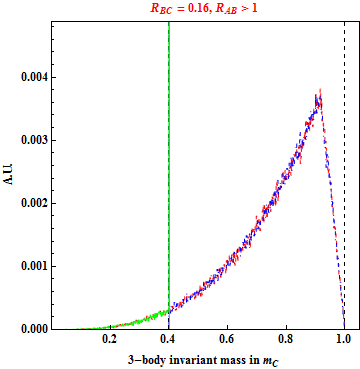} \\
\includegraphics[width=3.6cm]{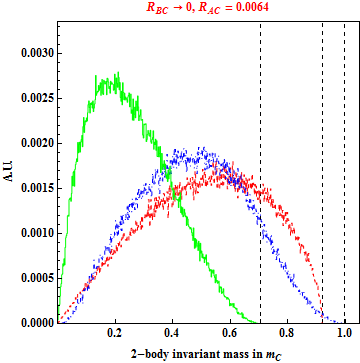}
\includegraphics[width=3.6cm]{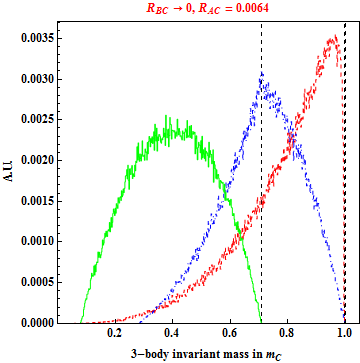}
\includegraphics[width=3.6cm]{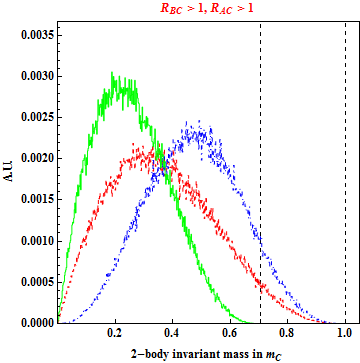}
\includegraphics[width=3.6cm]{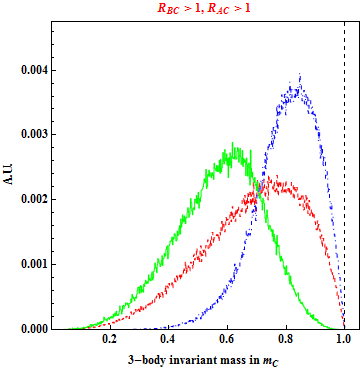} \\
\includegraphics[width=3.6cm]{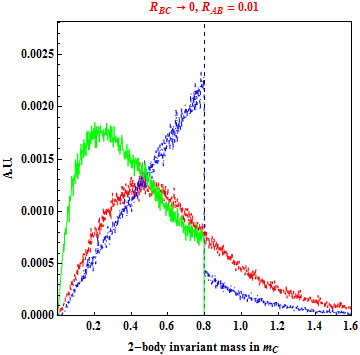}
\includegraphics[width=3.6cm]{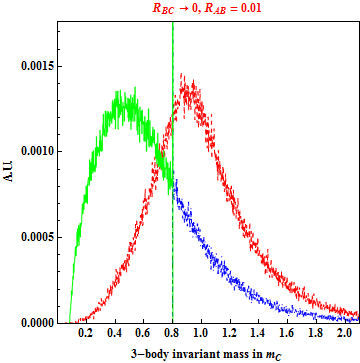}
\includegraphics[width=3.6cm]{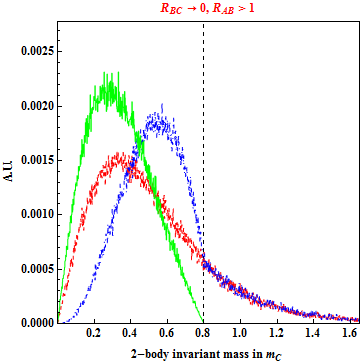}
\includegraphics[width=3.6cm]{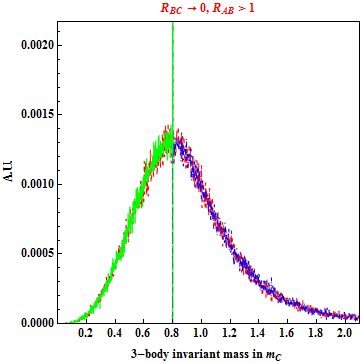}
\caption{\label{fig:theoryDist} Invariant mass distributions in regions $\mathcal{R}_{1(1)}$ (first two panels in the 1st row), $\mathcal{R}_{1(2)}$ (last two panels in the 1st row), $\mathcal{R}_{2(1)}$ (first two panels in the 2nd row), $\mathcal{R}_{2(2)}$ (last two panels in the 2nd row), $\mathcal{R}_{3}$ (first two panels in the 3rd row), $\mathcal{R}_{4}$ (last two panels in the 3rd row), $\mathcal{R}_{5}$ (first two panels in the 4th row), and $\mathcal{R}_{6}$ (last two panels in the 4th row). In each pair of panels, the first one demonstrates the 2-body invariant mass variables such as $m_{\ell\ell}$ (red dashed), $m_{\ell j}^>$ (blue dot-dashed), and $m_{\ell j}^<$ (green solid) while the second one demonstrates the 3-body invariant mass variables such as $m_{\ell\ell j}$ (red dashed), $m_{\ell jj}^>$ (blue dot-dashed), and $m_{\ell jj}^<$ (green solid). Black dashed lines denote the relevant theoretical kinematic endpoints formulated in Section~\ref{sec:derivation}.}
\end{figure}

We illustrate the invariant mass distributions for the regions discussed in Section~\ref{sec:derivation} in Figure~\ref{fig:theoryDist}. The relevant distributions for region $\mathcal{R}_7$ are not shown here because all kinematic endpoints (in squared mass) are simply given by either $s$ or $s/2$, hence less informative than the others. 
Among the eight possible invariant mass variables \eqref{eq:2body}-\eqref{eq:4body}, two are trivial, viz., $m_{\ell\ell jj}$ always peaks at $m_{W_R}$ and $m_{jj}$ peaks at $m_{W}$ for some cases; hence they are not considered here. For any event, the decay process of an on-shell resonance is performed through pure phase space. The events are generated with $\sqrt{s}=14$ TeV for all regions. As clearly mentioned in Section~\ref{sec:derivation}, the mass of particle $C$ merely determines the overall scale (for the on-shell $C$-initiated processes), while all the details are completely governed by the mass ratios $R_{ij}$. In this sense, the exact values of all mass parameters are irrelevant. Every distribution is plotted in the unit of $m_C$ which is fixed to be 1 TeV.

\begin{thebibliography}{99}
\bibitem{PDG} K. Nakamura and S. T. Petcov, Chapter 14 in K.~A.~Olive {\it et al.} [Particle Data Group Collaboration],
  Chin.\ Phys.\ C {\bf 38}, 090001 (2014), pp. 235-258. [{\tt http://pdg.lbl.gov/}]


\bibitem{Minkowski:1977sc}
  P.~Minkowski,
  Phys.\ Lett.\ B {\bf 67}, 421 (1977).

\bibitem{Mohapatra:1979ia}
  R.~N.~Mohapatra and G.~Senjanovi\'{c},
  Phys.\ Rev.\ Lett.\  {\bf 44}, 912 (1980).

\bibitem{Yanagida:1979as}
  T.~Yanagida,
  Conf.\ Proc.\ C {\bf 7902131}, 95 (1979).

\bibitem{GellMann:1980vs}
  M.~Gell-Mann, P.~Ramond and R.~Slansky,
  Conf.\ Proc.\ C {\bf 790927}, 315 (1979). 

\bibitem{Glashow:1979nm}
  S.~L.~Glashow,
  NATO Sci.\ Ser.\ B {\bf 61}, 687 (1980).

\bibitem{Mohapatra:2006gs} 
  R.~N.~Mohapatra and A.~Y.~Smirnov,
  Ann.\ Rev.\ Nucl.\ Part.\ Sci.\  {\bf 56}, 569 (2006)
  [hep-ph/0603118].

\bibitem{Malinsky:2005bi} 
  M.~Malinsky, J.~C.~Romao and J.~W.~F.~Valle,
  Phys.\ Rev.\ Lett.\  {\bf 95}, 161801 (2005)
  [hep-ph/0506296].

\bibitem{Dev:2009aw} 
  P.~S.~B.~Dev and R.~N.~Mohapatra,
  Phys.\ Rev.\ D {\bf 81}, 013001 (2010)
  [arXiv:0910.3924 [hep-ph]].  

\bibitem{Dev:2010he} 
  P.~S.~B.~Dev and R.~N.~Mohapatra,
  Phys.\ Rev.\ D {\bf 82}, 035014 (2010)
  [arXiv:1003.6102 [hep-ph]].

\bibitem{Borah:2010kk} 
  D.~Borah and U.~A.~Yajnik,
  Phys.\ Rev.\ D {\bf 83}, 095004 (2011)
  [arXiv:1010.6289 [hep-ph]].

\bibitem{Romeri:2011ie} 
  V.~De Romeri, M.~Hirsch and M.~Malinsky,
  Phys.\ Rev.\ D {\bf 84}, 053012 (2011)
  [arXiv:1107.3412 [hep-ph]].

\bibitem{Arbelaez:2013hr} 
  C.~Arbelaez, R.~M.~Fonseca, M.~Hirsch and J.~C.~Romao,
  Phys.\ Rev.\ D {\bf 87}, 075010 (2013)
  [arXiv:1301.6085 [hep-ph]].

\bibitem{Awasthi:2013ff} 
  R.~L.~Awasthi, M.~K.~Parida and S.~Patra,
  JHEP {\bf 1308}, 122 (2013)
  [arXiv:1302.0672 [hep-ph]].

 \bibitem{Deppisch:2014zta} 
  F.~F.~Deppisch, T.~E.~Gonzalo, S.~Patra, N.~Sahu and U.~Sarkar,
  Phys.\ Rev.\ D {\bf 91}, 015018 (2015)
  [arXiv:1410.6427 [hep-ph]].

\bibitem{Parida:2014dla} 
  M.~K.~Parida and B.~Sahoo,
  arXiv:1411.6748 [hep-ph].

\bibitem{Dev:2015pga} 
  P.~S.~B.~Dev and R.~N.~Mohapatra,
Phys.\ Rev.\ Lett.\  {\bf 115}, 181803 (2015)   
[arXiv:1508.02277 [hep-ph]].



\bibitem{Deppisch:2015cua} 
  F.~F.~Deppisch {\em et al.}, 
   Phys.\ Rev.\ D {\bf 93}, 013011 (2016) [arXiv:1508.05940 [hep-ph]].

\bibitem{Bandyopadhyay:2015fka} 
  T.~Bandyopadhyay, B.~Brahmachari and A.~Raychaudhuri,
  arXiv:1509.03232 [hep-ph].

\bibitem{Vissani:1997ys}
  F.~Vissani,
  Phys.\ Rev.\ D {\bf 57}, 7027 (1998)
  [hep-ph/9709409].

\bibitem{Clarke:2015hta}
  J.~D.~Clarke, R.~Foot and R.~R.~Volkas,
  Phys.\ Rev.\ D {\bf 91}, 073009 (2015)
  [arXiv:1502.01352 [hep-ph]].

\bibitem{Keung:1983uu} 
  W.~Y.~Keung and G.~Senjanovic,
  Phys.\ Rev.\ Lett.\  {\bf 50}, 1427 (1983).

\bibitem{Datta:1993nm} 
  A.~Datta, M.~Guchait and A.~Pilaftsis,
  Phys.\ Rev.\ D {\bf 50}, 3195 (1994)
  [hep-ph/9311257].



\bibitem{Dev:2013wba}
  P.~S.~B.~Dev, A.~Pilaftsis and U.~k.~Yang,
  Phys.\ Rev.\ Lett.\  {\bf 112}, 081801 (2014)
  [arXiv:1308.2209 [hep-ph]].




\bibitem{Alva:2014gxa}
  D.~Alva, T.~Han and R.~Ruiz,
  JHEP {\bf 1502}, 072 (2015)
  [arXiv:1411.7305 [hep-ph]].

\bibitem{Rodejohann:2011mu} 
  W.~Rodejohann,
  Int.\ J.\ Mod.\ Phys.\ E {\bf 20}, 1833 (2011)
  [arXiv:1106.1334 [hep-ph]].

\bibitem{deGouvea:2013zba} 
  A.~de Gouvea and P.~Vogel,
  Prog.\ Part.\ Nucl.\ Phys.\  {\bf 71}, 75 (2013)
  [arXiv:1303.4097 [hep-ph]].




\bibitem{Dev:2012zg} 
  P.~S.~B.~Dev, R.~Franceschini and R.~N.~Mohapatra,
  Phys.\ Rev.\ D {\bf 86}, 093010 (2012)
  [arXiv:1207.2756 [hep-ph]].

\bibitem{Cely:2012bz} 
  C.~G.~Cely, A.~Ibarra, E.~Molinaro and S.~T.~Petcov,
  Phys.\ Lett.\ B {\bf 718}, 957 (2013)
  [arXiv:1208.3654 [hep-ph]].

\bibitem{Maiezza:2015lza} 
  A.~Maiezza, M.~Nemevšek and F.~Nesti,
  Phys.\ Rev.\ Lett.\  {\bf 115}, 081802 (2015)
  [arXiv:1503.06834 [hep-ph]].

\bibitem{Dermisek:2015vra} 
  R.~Dermisek, E.~Lunghi and S.~Shin,
  JHEP {\bf 1508}, 126 (2015)
  [arXiv:1503.08829 [hep-ph]].

\bibitem{Atre:2009rg}
  A.~Atre, T.~Han, S.~Pascoli and B.~Zhang,
  JHEP {\bf 0905}, 030 (2009)
  [arXiv:0901.3589 [hep-ph]].

\bibitem{Drewes:2013gca} 
  M.~Drewes,
  Int.\ J.\ Mod.\ Phys.\ E {\bf 22}, 1330019 (2013)
  [arXiv:1303.6912 [hep-ph]].


\bibitem{Deppisch:2015qwa}
  F.~F.~Deppisch, P.~S.~B.~Dev and A.~Pilaftsis,
  New J.\ Phys.\  {\bf 17}, 075019 (2015)
  [arXiv:1502.06541 [hep-ph]].

\bibitem{Alekhin:2015byh} 
  S.~Alekhin {\it et al.},
  arXiv:1504.04855 [hep-ph].

\bibitem{Pati:1974yy} 
  J.~C.~Pati and A.~Salam,
  Phys.\ Rev.\ D {\bf 10}, 275 (1974)
  [Phys.\ Rev.\ D {\bf 11}, 703 (1975)].


\bibitem{Mohapatra:1974hk}
  R.~N.~Mohapatra and J.~C.~Pati,
  Phys.\ Rev.\ D {\bf 11}, 566 (1975).

\bibitem{Mohapatra:1974gc}
  R.~N.~Mohapatra and J.~C.~Pati,
  Phys.\ Rev.\ D {\bf 11}, 2558 (1975).

\bibitem{Senjanovic:1975rk}
  G.~Senjanovic and R.~N.~Mohapatra,
  Phys.\ Rev.\ D {\bf 12}, 1502 (1975).

\bibitem{Dev:2013oxa}
  C.-H.~Lee, P.~S.~B.~Dev and R.~N.~Mohapatra,
  Phys.\ Rev.\ D {\bf 88}, 093010 (2013)
  [arXiv:1309.0774 [hep-ph]].

\bibitem{Ferrari:2000sp} 
  A.~Ferrari {\em et al.},
  Phys.\ Rev.\ D {\bf 62}, 013001 (2000). 

\bibitem{Nemevsek:2011hz} 
  M.~Nemevsek, F.~Nesti, G.~Senjanovi\'{c} and Y.~Zhang,
  Phys.\ Rev.\ D {\bf 83}, 115014 (2011)
  [arXiv:1103.1627 [hep-ph]]. 

\bibitem{Chen:2011hc} 
  C.~Y.~Chen and P.~S.~B.~Dev,
  Phys.\ Rev.\ D {\bf 85}, 093018 (2012)
  [arXiv:1112.6419 [hep-ph]].

\bibitem{Chakrabortty:2012pp} 
  J.~Chakrabortty, J.~Gluza, R.~Sevillano and R.~Szafron,
  JHEP {\bf 1207}, 038 (2012)
  [arXiv:1204.0736 [hep-ph]]. 

\bibitem{Das:2012ii} 
  S.~P.~Das, F.~F.~Deppisch, O.~Kittel and J.~W.~F.~Valle,
  Phys.\ Rev.\ D {\bf 86}, 055006 (2012)
  [arXiv:1206.0256 [hep-ph]]. 

\bibitem{Saavedra:2012gf} 
  J.~A.~Aguilar-Saavedra and F.~R.~Joaquim,
  Phys.\ Rev.\ D {\bf 86}, 073005 (2012)
  [arXiv:1207.4193 [hep-ph]].

\bibitem{Chen:2013fna} 
  C.~Y.~Chen, P.~S.~B.~Dev and R.~N.~Mohapatra,
  Phys.\ Rev.\ D {\bf 88}, 033014 (2013)
  [arXiv:1306.2342 [hep-ph]].

\bibitem{Rizzo:2014xma} 
  T.~G.~Rizzo,
  Phys.\ Rev.\ D {\bf 89}, 095022 (2014)
  [arXiv:1403.5465 [hep-ph]].

\bibitem{Ng:2015hba} 
  J.~N.~Ng, A.~de la Puente and B.~W.~P.~Pan,
  JHEP {\bf 1512}, 172 (2015) [arXiv:1505.01934 [hep-ph]].


\bibitem{Mohapatra:1981pm} 
  R.~N.~Mohapatra and J.~D.~Vergados,
  Phys.\ Rev.\ Lett.\  {\bf 47}, 1713 (1981).

\bibitem{Picciotto:1982qe} 
  C.~E.~Picciotto and M.~S.~Zahir,
  Phys.\ Rev.\ D {\bf 26}, 2320 (1982).

\bibitem{Hirsch:1996qw} 
  M.~Hirsch, H.~V.~Klapdor-Kleingrothaus and O.~Panella,
  Phys.\ Lett.\ B {\bf 374}, 7 (1996)
  [hep-ph/9602306].

\bibitem{Tello:2010am} 
  V.~Tello, M.~Nemevsek, F.~Nesti, G.~Senjanovic and F.~Vissani,
  Phys.\ Rev.\ Lett.\  {\bf 106}, 151801 (2011)
  [arXiv:1011.3522 [hep-ph]].

\bibitem{Chakrabortty:2012mh} 
  J.~Chakrabortty, H.~Z.~Devi, S.~Goswami and S.~Patra,
  JHEP {\bf 1208}, 008 (2012)
  [arXiv:1204.2527 [hep-ph]]. 

\bibitem{Nemevsek:2012iq} 
  M.~Nemevsek, G.~Senjanovic and V.~Tello,
  Phys.\ Rev.\ Lett.\  {\bf 110}, 151802 (2013)
  [arXiv:1211.2837 [hep-ph]].

\bibitem{Barry:2013xxa} 
J.~Barry and W.~Rodejohann,
  JHEP {\bf 1309}, 153 (2013) [arXiv:1303.6324 [hep-ph]]. 

\bibitem{Dev:2013vxa} 
  P.~S.~B. Dev, S.~Goswami, M.~Mitra and W.~Rodejohann,
  Phys.\ Rev.\ D {\bf 88}, 091301 (2013)
  [arXiv:1305.0056 [hep-ph]]. 

\bibitem{Huang:2013kma} 
  W.~C.~Huang and J.~Lopez-Pavon,
  Eur.\ Phys.\ J.\ C {\bf 74}, 2853 (2014)
  [arXiv:1310.0265 [hep-ph]].


\bibitem{Dev:2014xea} 
  P.~S.~B.~Dev, S.~Goswami and M.~Mitra,
  Phys.\ Rev.\ D {\bf 91}, 113004 (2015)
  [arXiv:1405.1399 [hep-ph]].

\bibitem{Ge:2015yqa} 
  S.~F.~Ge, M.~Lindner and S.~Patra,
  JHEP {\bf 1510}, 077 (2015) [arXiv:1508.07286 [hep-ph]].

\bibitem{Borah:2015ufa} 
  D.~Borah and A.~Dasgupta,
  JHEP {\bf 1511}, 208 (2015) [arXiv:1509.01800 [hep-ph]].


\bibitem{Awasthi:2015ota} 
  R.~L.~Awasthi, P.~S.~B.~Dev and M.~Mitra,
  arXiv:1509.05387 [hep-ph].

\bibitem{Riazuddin:1981hz} 
  Riazuddin, R.~E.~Marshak and R.~N.~Mohapatra,
  Phys.\ Rev.\ D {\bf 24}, 1310 (1981).

\bibitem{Pal:1983bf} 
  P.~B.~Pal,
  Nucl.\ Phys.\ B {\bf 227}, 237 (1983).

\bibitem{Mohapatra:1992uu} 
  R.~N.~Mohapatra,
  Phys.\ Rev.\ D {\bf 46}, 2990 (1992).
 

\bibitem{Cirigliano:2004mv} 
  V.~Cirigliano, A.~Kurylov, M.~J.~Ramsey-Musolf and P.~Vogel,
  Phys.\ Rev.\ D {\bf 70}, 075007 (2004)
  [hep-ph/0404233].

\bibitem{Cirigliano:2004tc} 
  V.~Cirigliano, A.~Kurylov, M.~J.~Ramsey-Musolf and P.~Vogel,
  Phys.\ Rev.\ Lett.\  {\bf 93}, 231802 (2004)
  [hep-ph/0406199].

\bibitem{Mohapatra:1986pj} 
  R.~N.~Mohapatra,
  Phys.\ Rev.\ D {\bf 34}, 909 (1986).


\bibitem{Helo:2013esa} 
  J.~C.~Helo, M.~Hirsch and S.~Kovalenko,
  Phys.\ Rev.\ D {\bf 89}, 073005 (2014)
  [arXiv:1312.2900 [hep-ph]].



\bibitem{Cvetic:2015naa} 
  G.~Cvetic, C.~Dib, C.~S.~Kim and J.~Zamora-Saa,
  Symmetry {\bf 7}, 726 (2015)
  [arXiv:1503.01358 [hep-ph]].

\bibitem{Felisola:2015bha} 
  O.~Castillo-Felisola, C.~O.~Dib, J.~C.~Helo, S.~G.~Kovalenko and S.~E.~Ortiz,
  Phys.\ Rev.\ D {\bf 92}, 013001 (2015)
  [arXiv:1504.02489 [hep-ph]].

\bibitem{Dib:2015oka} 
  C.~O.~Dib and C.~S.~Kim,
  Phys.\ Rev.\ D {\bf 92}, 093009 (2015) [arXiv:1509.05981 [hep-ph]].





\bibitem{Kersten:2007vk} 
  J.~Kersten and A.~Y.~Smirnov,
  Phys.\ Rev.\ D {\bf 76}, 073005 (2007)
  [arXiv:0705.3221 [hep-ph]].

\bibitem{Ibarra:2010xw} 
  A.~Ibarra, E.~Molinaro and S.~T.~Petcov,
  JHEP {\bf 1009}, 108 (2010)
  [arXiv:1007.2378 [hep-ph]].

\bibitem{Pavon:2015cga} 
  J.~Lopez-Pavon, E.~Molinaro and S.~T.~Petcov,
  arXiv:1506.05296 [hep-ph].


\bibitem{Akhmedov:2007fk} 
  E.~K.~Akhmedov,
  JHEP {\bf 0709}, 116 (2007)
  [arXiv:0706.1216 [hep-ph]].


\bibitem{Pilaftsis:1991ug} 
  A.~Pilaftsis,
  Z.\ Phys.\ C {\bf 55}, 275 (1992)
  [hep-ph/9901206].

\bibitem{Panella:2001wq} 
  O.~Panella, M.~Cannoni, C.~Carimalo and Y.~N.~Srivastava,
  Phys.\ Rev.\ D {\bf 65}, 035005 (2002)
  [hep-ph/0107308].

  \bibitem{Han:2006ip} 
  T.~Han and B.~Zhang,
  Phys.\ Rev.\ Lett.\  {\bf 97}, 171804 (2006)
  [hep-ph/0604064]. 

\bibitem{Aguila:2007em} 
  F.~del Aguila, J.~A.~Aguilar-Saavedra and R.~Pittau,
  JHEP {\bf 0710}, 047 (2007)
  [hep-ph/0703261].

\bibitem{Aguila:2008cj} 
  F.~del Aguila and J.~A.~Aguilar-Saavedra,
  Nucl.\ Phys.\ B {\bf 813}, 22 (2009)
  [arXiv:0808.2468 [hep-ph]].

\bibitem{Mohapatra:1986aw} 
  R.~N.~Mohapatra,
  Phys.\ Rev.\ Lett.\  {\bf 56}, 561 (1986).

\bibitem{Mohapatra:1986bd} 
  R.~N.~Mohapatra and J.~W.~F.~Valle,
  Phys.\ Rev.\ D {\bf 34}, 1642 (1986).

\bibitem{Akhmedov:1995ip} 
  E.~K.~Akhmedov, M.~Lindner, E.~Schnapka and J.~W.~F.~Valle,
  Phys.\ Lett.\ B {\bf 368}, 270 (1996)
  [hep-ph/9507275].

\bibitem{Barr:2003nn} 
  S.~M.~Barr,
  Phys.\ Rev.\ Lett.\  {\bf 92}, 101601 (2004)
  [hep-ph/0309152].


\bibitem{Gavela:2009cd} 
  M.~B.~Gavela, T.~Hambye, D.~Hernandez and P.~Hernandez,
  JHEP {\bf 0909}, 038 (2009)
  [arXiv:0906.1461 [hep-ph]].

\bibitem{Dev:2012sg} 
  P.~S.~B.~Dev and A.~Pilaftsis,
  Phys.\ Rev.\ D {\bf 86}, 113001 (2012)
  [arXiv:1209.4051 [hep-ph]].

\bibitem{Khachatryan:2014dka} 
  V.~Khachatryan {\it et al.} [CMS Collaboration],
  Eur.\ Phys.\ J.\ C {\bf 74}, 3149 (2014)
  [arXiv:1407.3683 [hep-ex]].

\bibitem{Aad:2015xaa} 
  G.~Aad {\it et al.} [ATLAS Collaboration],
  JHEP {\bf 1507}, 162 (2015)
  [arXiv:1506.06020 [hep-ex]].

\bibitem{Gluza:2015goa}
  J.~Gluza and T.~Jeli\'{n}ski,
  Phys.\ Lett.\ B {\bf 748}, 125 (2015)
  [arXiv:1504.05568 [hep-ph]].

\bibitem{Dobrescu:2015qna}  B.~A.~Dobrescu and Z.~Liu, 
  JHEP {\bf 1510}, 118 (2015) [arXiv:1507.01923 [hep-ph]].


\bibitem{Dev:2013vba} 
  P.~S.~B.~Dev and R.~N.~Mohapatra,
  arXiv:1308.2151 [hep-ph].


\bibitem{Miller:2005zp} 
  D.~J.~Miller, P.~Osland and A.~R.~Raklev,
  JHEP {\bf 0603}, 034 (2006)
  [hep-ph/0510356].


\bibitem{Agashe:2010gt}
  K.~Agashe, D.~Kim, M.~Toharia and D.~G.~E.~Walker,
  Phys.\ Rev.\ D {\bf 82}, 015007 (2010)
  [arXiv:1003.0899 [hep-ph]].

\bibitem{Agashe:2010tu}
  K.~Agashe, D.~Kim, D.~G.~E.~Walker and L.~Zhu,
  Phys.\ Rev.\ D {\bf 84}, 055020 (2011)
  [arXiv:1012.4460 [hep-ph]].

\bibitem{Cho:2012er}
  W.~S.~Cho, D.~Kim, K.~T.~Matchev and M.~Park,
  Phys.\ Rev.\ Lett.\  {\bf 112}, 211801 (2014)
  [arXiv:1206.1546 [hep-ph]].

\bibitem{Agashe:2012fs}
  K.~Agashe, R.~Franceschini, D.~Kim and K.~Wardlow,
  Phys.\ Dark Univ.\  {\bf 2}, 72 (2013)
  [arXiv:1212.5230 [hep-ph]].


\bibitem{Han:2012vk} 
  T.~Han, I.~Lewis, R.~Ruiz and Z.~g.~Si,
  Phys.\ Rev.\ D {\bf 87}, 035011 (2013)
  [Erratum {\em ibid}, 039906 (2013)]
  [arXiv:1211.6447 [hep-ph]].

\bibitem{Beall:1981ze} 
  G.~Beall, M.~Bander and A.~Soni,
  Phys.\ Rev.\ Lett.\  {\bf 48}, 848 (1982).

\bibitem{Zhang:2007da} 
  Y.~Zhang, H.~An, X.~Ji and R.~N.~Mohapatra,
  Nucl.\ Phys.\ B {\bf 802}, 247 (2008)
  [arXiv:0712.4218 [hep-ph]].

\bibitem{Maiezza:2010ic} 
  A.~Maiezza, M.~Nemevsek, F.~Nesti and G.~Senjanovic,
  Phys.\ Rev.\ D {\bf 82}, 055022 (2010)
  [arXiv:1005.5160 [hep-ph]].

\bibitem{Bertolini:2014sua} 
  S.~Bertolini, A.~Maiezza and F.~Nesti,
  Phys.\ Rev.\ D {\bf 89}, 095028 (2014)
  [arXiv:1403.7112 [hep-ph]].


\bibitem{Burns:2009zi}
  M.~Burns, K.~T.~Matchev and M.~Park,
  JHEP {\bf 0905}, 094 (2009)
  [arXiv:0903.4371 [hep-ph]].

\bibitem{Lester:2006cf}
  C.~G.~Lester, M.~A.~Parker and M.~J.~White,
  JHEP {\bf 0710}, 051 (2007)
  [hep-ph/0609298].

\bibitem{KMP}
  D.~Kim, K.~Matchev and M.~Park, 
arXiv:1512.02222 [hep-ph].
  
\bibitem{Helo:2013dla} 
  J.~C.~Helo, M.~Hirsch, S.~G.~Kovalenko and H.~P\"{a}s,
  Phys.\ Rev.\ D {\bf 88}, no. 1, 011901 (2013)
  [arXiv:1303.0899 [hep-ph]].
  
\bibitem{Helo:2013ika} 
  J.~C.~Helo, M.~Hirsch, H.~P\"{a}s and S.~G.~Kovalenko,
  Phys.\ Rev.\ D {\bf 88}, 073011 (2013)
  [arXiv:1307.4849 [hep-ph]].


\bibitem{Alwall:2014hca}
  J.~Alwall {\it et al.},
  JHEP {\bf 1407}, 079 (2014)
  [arXiv:1405.0301 [hep-ph]].

\bibitem{Ball:2012cx}
  R.~D.~Ball  {\it et al.},
  Nucl.\ Phys.\ B {\bf 867}, 244 (2013)
  [arXiv:1207.1303 [hep-ph]].

\bibitem{Sjostrand:2006za}
  T.~Sjostrand, S.~Mrenna and P.~Z.~Skands,
  JHEP {\bf 0605}, 026 (2006)
  [hep-ph/0603175].

\bibitem{deFavereau:2013fsa}
  J.~de Favereau {\it et al.}  [DELPHES 3 Collaboration],
  JHEP {\bf 1402}, 057 (2014)
  [arXiv:1307.6346 [hep-ex]].

\bibitem{Cacciari:2011ma}
  M.~Cacciari, G.~P.~Salam and G.~Soyez,
  Eur.\ Phys.\ J.\ C {\bf 72}, 1896 (2012)
  [arXiv:1111.6097 [hep-ph]].

\bibitem{Cacciari:2008gp}
  M.~Cacciari, G.~P.~Salam and G.~Soyez,
  JHEP {\bf 0804}, 063 (2008)
  [arXiv:0802.1189 [hep-ph]].

\bibitem{Dokshitzer:1997in}
  Y.~L.~Dokshitzer, G.~D.~Leder, S.~Moretti and B.~R.~Webber,
  JHEP {\bf 9708}, 001 (1997)
  [hep-ph/9707323].

\bibitem{Wobisch:1998wt}
  M.~Wobisch and T.~Wengler,
  in {\em Hamburg 1998/1999: Monte Carlo generators for HERA physics},  pp. 270-279
  [hep-ph/9907280].

\bibitem{Butterworth:2008iy}
  J.~M.~Butterworth, A.~R.~Davison, M.~Rubin and G.~P.~Salam,
  Phys.\ Rev.\ Lett.\  {\bf 100}, 242001 (2008)
  [arXiv:0802.2470 [hep-ph]].

\bibitem{Kim:2014ana}
  D.~Kim, H.~S.~Lee and M.~Park,
  JHEP {\bf 1503}, 134 (2015)
  [arXiv:1411.0668 [hep-ph]].

\bibitem{Khachatryan:2015gha} 
  V.~Khachatryan {\it et al.} [CMS Collaboration],
  Phys.\ Lett.\ B {\bf 748}, 144 (2015)
  [arXiv:1501.05566 [hep-ex]].



\bibitem{Antusch:2014woa} 
  S.~Antusch and O.~Fischer,
  JHEP {\bf 1410}, 94 (2014)
  [arXiv:1407.6607 [hep-ph]].





\bibitem{Agostini:2013mzu}
M.~Agostini {\it et al.} [GERDA Collaboration],
  Phys.\ Rev.\ Lett.\  {\bf 111}, 122503 (2013) [arXiv:1307.4720 [nucl-ex]].

\bibitem{Meroni:2012qf} 
  A.~Meroni, S.~T.~Petcov and F.~Simkovic,
  JHEP {\bf 1302}, 025 (2013)
  [arXiv:1212.1331 [hep-ph]].

\bibitem{Abgrall:2013rze} 
  N.~Abgrall {\it et al.} [Majorana Collaboration],
  Adv.\ High Energy Phys.\  {\bf 2014}, 365432 (2014)
  [arXiv:1308.1633 [physics.ins-det]].

\bibitem{Arnold:2010tu} 
  R.~Arnold {\it et al.} [SuperNEMO Collaboration],
  Eur.\ Phys.\ J.\ C {\bf 70}, 927 (2010)
  [arXiv:1005.1241 [hep-ex]].

\bibitem{Barry:2012ga} 
  J.~Barry, L.~Dorame and W.~Rodejohann,
  Eur.\ Phys.\ J.\ C {\bf 72}, 2023 (2012)
  [arXiv:1203.3365 [hep-ph]].

\bibitem{Adam:2013mnn} 
  J.~Adam {\it et al.} [MEG Collaboration],
  Phys.\ Rev.\ Lett.\  {\bf 110}, 201801 (2013)
  [arXiv:1303.0754 [hep-ex]].

\bibitem{Baldini:2013ke} 
  A.~M.~Baldini {\it et al.} [MEG Collaboration],
  arXiv:1301.7225 [physics.ins-det]. 

\bibitem{Hinchliffe:2015qma} 
  I.~Hinchliffe, A.~Kotwal, M.~L.~Mangano, C.~Quigg and L.~T.~Wang,
  Int.\ J.\ Mod.\ Phys.\ A {\bf 30}, 1544002 (2015)
  [arXiv:1504.06108 [hep-ph]].

\bibitem{Frere:2008ct} 
  J.~M.~Frere, T.~Hambye and G.~Vertongen,
  JHEP {\bf 0901}, 051 (2009)
  [arXiv:0806.0841 [hep-ph]].

\bibitem{Dev:2014iva} 
  P.~S.~B. Dev, C.~H.~Lee and R.~N.~Mohapatra,
  Phys.\ Rev.\ D {\bf 90}, 095012 (2014)
  [arXiv:1408.2820 [hep-ph]]; 
J.\ Phys.\ Conf.\ Ser.\  {\bf 631}, 012007 (2015)
  [arXiv:1503.04970 [hep-ph]].

\end{thebibliography}
\end{document}